\batchmode
\documentclass[12pt]{article}    
\usepackage{hyperref}
\usepackage{graphics}
\usepackage{epsfig}
\usepackage{natbib}   
\usepackage{graphicx} 
\usepackage{amssymb}  
\usepackage{multirow} 
\usepackage{epstopdf}
\usepackage{bm}
\usepackage{amsmath}
\usepackage{amsfonts}
\usepackage{amssymb}
\usepackage{pdflscape}
\usepackage{color}
\usepackage{epstopdf}
\usepackage{rotating}
\usepackage{array}
\usepackage{longtable}
\usepackage{lmodern}

\def \R{{\rm I\kern -2.2pt  R\hskip 1pt}}
\def \N{{\rm I\kern -2.2pt  N\hskip 1pt}}

\def\eu{\,\hbox{\raise .36em\hbox to0pt{\vrule height0.5pt%
width.55em depth0pt\hss}%
\raise .25em\hbox to0pt{\vrule height0.5pt width.5em%
depth0pt\hss}\hskip.02em\sf C}\,}
\begin{document}
\newtheorem{definition}{Definition}
\newtheorem{theorem}{Theorem}
\newtheorem{example}{Example}
\newtheorem{corollary}{Corollary}
\newtheorem{lemma}{Lemma}
\newtheorem{proposition}{Proposition}
\newenvironment{proof}{{\bf Proof:\ \ }}{\qed}
\newcommand{\qed}{\rule{0.5em}{1.5ex}}
\newcommand{\bfg}[1]{\mbox{\boldmath $#1$\unboldmath}}
\vspace*{-4cm}

\begin{center}

\section*{Modeling the obsolescence of research literature in disciplinary journals through the age of their cited references}

\vskip 0.2in {\sc \bf Pablo Dorta--Gonz\'alez, Emilio G\'omez--D\'eniz  }

\vskip 0.1in

Department of Quantitative Methods and TIDES Institute,
University of Las Palmas de Gran Canaria, Campus de Tafira s/n,
35017--Las Palmas de Gran Canaria, Spain,\\{\tt pablo.dorta@ulpgc.es}\\ {\tt emilio.gomez-deniz@ulpgc.es}\\
http://orcid.org/0000-0003-0494-2903(PDG)\\
http://orcid.org/0000-0002-5072-7908 (EGD)

\end{center}

\begin{abstract}\noindent There are different citation habits in the research fields that influence the obsolescence of the research literature. We analyze the distinctive obsolescence of {research literature in disciplinary journals in} eight scientific subfields based on cited references distribution, as a synchronous approach. We use both Negative Binomial (NB) and Poisson distributions to capture this obsolescence. The corpus being examined is published in 2019 and covers 22,559 papers citing 872,442 references. Moreover, three measures to analyze the tail of the distribution are proposed: {(i) cited reference survival rate, (ii) cited reference mortality rate, and (iii) cited reference percentile.} These measures are interesting because the tail of the distribution collects the behavior of the citations at the time when the document starts to get obsolete in the sense that it is little cited (used). As main conclusion, the differences observed in obsolescence are so important even between {disciplinary} journals in the same subfield, that it would be necessary to use some measure for the tail of the citation distribution, such as those proposed in this paper, when analyzing in an appropriate way the {long time} impact of a journal.

\end{abstract}

\noindent \textit{Key Words}: literature obsolescence, literature aging, {long time impact}, citation distribution, Negative Binomial distribution, Poisson distribution, citation half-life, citation life expectancy

\medskip

\noindent {\bf Mathematics Subject Classification 2020:} 62E10, 62P25.\vfill

\subsubsection*{Funding:}
Emilio G\'omez-D\'eniz was partially funded by grant
ECO2017--85577--P (Ministerio de Econom\'ia, Industria y
Competitividad. Agencia Estatal de Investigaci\'on).

{\subsubsection*{Acknowledgements}
We thank the two anonymous reviewers for their valuable comments and suggestions, which have greatly helped us improve the original manuscript.}

\section{Introduction}The use of research publications decreases with time and age of literature. In library and information science, this phenomenon is known as literature obsolescence.  This refers to a decrease in its frequency of use or citation, but not to its value as a source of knowledge.

The obsolescence in the research literature can be measure through the citation analysis technique. This methodology supposes a relationship between the cited document and the citing one. Authors increasingly cite current documents in detriment of older, which can be consider obsolete because they stop being cited (used).

Studies on modelling literature obsolescence go back to the sixties of last century. \cite{DeSollaPrice1965} postulated that the use of literature declines with time according to a negative exponential distribution, although other authors suggested a lognormal distribution as the most suitable to measure literature obsolescence \citetext{\citealp{Egghe1992}; \citealp{Gupta1998}}.

There are two methods to study obsolescence: synchronic (retrospective) and diachronic (prospective). Synchronic analysis is based on references made, while diachronic analysis is based on citations received. In synchronic obsolescence analysis the half-life is the statistical median considering the years of the references in reverse chronological order. Diachronic obsolescence requires setting a period of time in the past to look into the future. Both methods can also be conducted retrospectively \citetext{\citealp{Egghe2000}}. However, some authors argue that synchronic and diachronic studies produce similar results \citetext{\citealp{Stinson1987}}, suggesting a preference for the synchronic method.

Although it is possible to estimate literature's half-life and even the time when the literature is not cited anymore, it is not possible to know the reasons for this. The main factor pointed out in the literature to justify the different obsolescence patterns is the diffusion process of knowledge in the different fields. {Two main types of obsolescence occur related to the diffusion process. Both types are related to the rate at which diffusion occurs. In the first place, initially, there is a high number of citations until a modal value is reached, followed by a rapid drop in them, a drop that is also very sharp, as occurs, for example, in the fields of Medicine and Engineering. The second type is due to a slower diffusion process, associated with a slower rate of decline. Examples of this behavior in obsolescence are the basic sciences, social sciences, and humanities} \citetext{\citealp{Bouabid2013}; \citealp{canoandlind_1991}}.

The objective of this work is to measure {the obsolescence of research literature in disciplinary journals}. We present data and results on the obsolescence of {eighty disciplinary journals over a long time period.}

We analyze the distinctive obsolescence of {research literature in disciplinary journals in} eight scientific subfields based on cited references distribution, as a synchronous approach. We use both Negative Binomial (NB) and Poisson distributions to capture this obsolescence. The corpus being examined is published in 2019 and covers 22,559 papers citing 872,442 references. Moreover, three measures to analyze the tail of the distribution are proposed: {(i) survival rate, (ii) mortality rate, and (iii)  percentile.}

These measures of the distribution tail collect the behavior of the citations at the time when the document starts to get obsolete in the sense that it is little cited (used). The theoretical framework is described in the following section.

\section{Theoretical framework in obsolescence of the research literature}
Since \cite{Gross1927} introduced the concept of obsolescence, that is, the phenomenon that research publications are decreasingly used over time, obsolescence and its law have become an important topic in bibliometrics and scientometrics.

In library and information sciences, the term half-life first appeared in the work by \cite{Burton1960}. These authors postulated that literature becomes obsolete and half-life means the time during which one-half of the currently active literature was published.

The use of literature declines with time according to a negative exponential distribution \citetext{\citealp{DeSollaPrice1965}}. In this respect, \cite{Ewing1966} carried out a diachronic study of research articles in chemistry and found that the number of citations decreased as the year of publication was closer to the current year. This author pointed to the growth of the literature as a factor that influences the measurement of the obsolescence ratio.

Disambiguating the term half-life, \cite{Line1970} asserts that literature half-life should be composed by the obsolescence rate and the literature growth rate. \cite{Line1974} defined obsolescence as the decrease or fall into time of the validity or utility of information. However, \cite{Brookes1970} stated that the theoretical problem of measuring obsolescence rate when literature is growing is more complex that the method proposed by Line.

Citation density decreases exponentially with age \citetext{\citealp{ Gupta1990}}. However, literature obsolescence can also be influenced by unknown factors. Thus, \cite{Egghe1992} demonstrated that the obsolescence factor defined by \cite{Brookes1970} is not actually a constant, but a statistical function of time. This is because citation data is not exponentially distributed as Brookes stated. In general, citation distribution presents an initial growth followed by an exponential decline. These authors stated that lognormal distribution is the model that best describes both the initial growth of citations and the subsequent decline.

About the combination of obsolescence and growth, both phenomena can be studied with the same mathematical function \citetext{\citealp{ Egghe1992}}. In the synchronic case, obsolescence increases with literature growth; in the diachronic case, the effect is opposite. In this respect, \cite{VanRaan2000} argues that in the initial moments of any discipline much less published documents exist than in later years. Thus, citation distribution according to years always has a combination of aging and literature growth phenomena.

\cite{Burrell2002} concluded that the lognormal distribution model shows more success describing the synchronic citations of documents, confirming previous studies such as that of \cite{Egghe1992}. These authors also observed that lognormal distribution describes and fits very well to the age of the first references \citetext{\citealp{Egghe2002}}.

Some studies compare obsolescence patterns across disciplines \citetext{\citealp{lariviereandarchambault_2008}; \citealp{songmaandyang_2015}; \citealp{Zhang_2017,Zhangetal2017}}. The main factor pointed out in the literature to justify the different obsolescence patterns is the diffusion process of knowledge \citetext{\citealp{Bouabid2013}; \citealp{canoandlind_1991}; \citealp{smallandcrane_1979}}. Some authors developed a mathematical model to describe this pattern \citetext{\citealp{Barnettetal1989}}.

An overview of the literature aging is offered by \cite{glanzel_2004}, presenting the different aspects that can be analyzed by both methods, synchronic and diachronic. More recently, studies and discussions on this matter continued with aspects as the mathematical model \citetext{\citealp{Bouabid2011}; \citealp{Wallace2009}}, obsolescence measures \citetext{\citealp{Bouabid2013}, \citealp{Zhang_2017,Zhangetal2017}}, and its influencing factors \citetext{\citealp{Wang2019}}.

\section{Empirical data}
The Scopus database was used as data source for the empirical application. We considered the following eight subject categories (subfields) in this database, three of them from science, two from social science, one from health science, one from engineering, and one from humanities: Cell Biology, Economics \& Econometrics, Electrical \& Electronic Engineering, General Chemistry, General Medicine, General Physics \& Astronomy, History, and Library \& Information Sciences.

We decided a priori to take eight subfields. This number was set so that both figures and tables could be displayed in the paper. The subfields were selected based on the previous experience of the authors and trying to cover disciplines as diverse as possible.

For each subfield, ten {disciplinary journals} were randomly selected from the top 10\% of most cited journals in the subfield measured through the Scopus CiteScore. Note that there is great variability in the number of journals per subfields (size of the subject categories). Some of them are very large, with more than 300 journals. For this reason, only ten journals were randomly selected from the top 10\% in relation to the CiteScore.

Then, for each journal, all documents published in 2019 and catalogued in this database as research articles were collected. Finally, for each research article, the years of publication from all cited references under the age of 150 were downloaded.

All the journals used here and its abbreviated title together with its corresponding subject category are shown in Table \ref{tabA1} (see the Appendix).

\section{Methodology}
Most of the studies in aging of cited references are based on the fact that the distribution of age for the cited references is a continuous
distribution. The exponential distribution is considered the first assumption as a simple case of working. In this case \citetext{see for
instance \citealp{Egghe2002}} is considered that the number of citations received at time $t$ is $c(t)=\theta\exp(-\theta t)$, $t\geq 0$, $\theta>0$, for which the obsolescence (aging) factor $a$ results
$a=c(t+1)/c(t)=\exp(-\theta)$ which is independent of $t$.

Here, we do not use the concept of obsolescence distribution function and will consider that $X_1,X_2,\dots,X_t$ are independent and identically distributed random variables with values in ${\cal X}\in{\mathbb N}$. They represent the number of cited references in a journal or a collection of journals in a subject category in the last $t$ years. That is, for a given sample of documents, $X_i$ represents the number of cited references with an age of $i$ years. Following to \cite{Burrell2002} we are going to assume that $X\equiv X_i$ initially follows a Poisson distribution with mean $\theta\in\Theta>0$. That is,
{\begin{eqnarray}
\Pr(X =x|\theta)=\frac{\theta ^{x}}{x!}\exp(-\theta),\quad x=0,1,\dots\label{poisson}
\end{eqnarray}}

In practise the citation of papers empirically seems to decrease with the years and this behaviour can be modelled by the Poisson distribution. Nevertheless, the Poisson distribution shows equi-dispersion (i.e. the variance is equal to the mean), which would make it inappropriate for defining the random variable $X$, an event that empirical studies have shown presents over-dispersion (i.e. the variance is greater than the mean). See Table \ref{tab1} where the index of dispersion ($\mbox{ID}=var(X)/E(X)$) is shown together with other descriptive statistics of the empirical data used here.
\begin{table}[htbp]
\caption{Some descriptive statistics of the subject categories empirical data\label{tab1}}
\begin{center}
\begin{tabular}{lcccccccc}\hline
& CB & E\&E & E\&EE & GC & GM  & GP\&A & H & L\&IS\\ \cline{2-9}
Mean & 9.63 & 12.72 & 7.92 & 8.98 & 8.57 & 14.10 & 21.72 & 11.49 \\
Median & 7 & 9 & 5 & 5 & 6 & 9 & 13 & 8\\
Mode & 2 & 3 & 2 & 2 & 2 & 2 & 3 & 3 \\
$\min$ & 0 & 0 & 0 & 0 & 0 & 0 & 0 & 0 \\
$\max$ &  147 & 148 & 144 & 150 & 149 & 150 & 150 & 146  \\
ID & 8.29 & 12.39 & 11.98 & 13.11 & 8.84 & 16.00 & 26.78 & 13.07\\ \hline
\end{tabular}
\end{center}
\end{table}

In addition to the above, there is evidence that over-dispersion is related to the heterogeneity of the population of subject
categories. In this case, the parameter $\theta$ can be considered a random variable that takes different values between
different journals in the subject category, reflecting uncertainty about this parameter and
varying from individual to individual according to a probability density function.  Here, we
assume that this parameter follows a gamma distribution with shaper parameter $\alpha>0$ and rate
parameter $\beta>0$, i.e.
\begin{eqnarray}
\pi(\theta) = \frac{\beta^{\alpha}}{\Gamma(\alpha)} \theta^{\alpha-1}\exp(-\beta\theta),\quad \theta>0.\label{gamma}
\end{eqnarray}

In practise, other mixing distribution beyond the gamma, as the inverse Gaussian distribution, could be considered here. Thus, {after using \eqref{poisson} and \eqref{gamma} we get that the unconditional distribution $\Pr(X=x)=\int_{\Theta}\Pr(X=x|\theta)\pi(\theta)\,d\theta$}
 results a negative binomial distribution, $X_t\sim NB(\alpha,1/(\beta+1))$. In this case, the obsolescence factor is given by
 $a(t)=\beta/(\beta+1)(1+(\alpha-1)/(t+1))$. It is well-known that for mixed Poisson distributions the variance is always greater than the mean.

\subsection{Cited reference survival rate and cited reference mortality rate}
The survival function describes the proportion of cited documents beyond a given value $x$, thus the probability that the document will survive beyond $x$. Survival is understood here in the sense that the document continues to be cited, if it has been until reaching the number of citations $x$. For the negative binomial distribution a closed-form expression for the survival
function is written in terms of the regularized generalized incomplete beta function, $I_z(a,b,c)$ \citetext{see \citealp[p. 45]{johnsonetal2005}} and
is given by,
\begin{eqnarray*}
R(X)=\Pr(X\geq x)=I_{\beta/(\beta+1)}(1,\alpha,x).
\end{eqnarray*}

The mortality function (also known as hazard function) is defined as
{$\lambda(x)=\Pr(X=x)/R(x)$}. In our setting, represents the probability, per unit of $x$, that the document ceases to be cited just after reaching the number of citations $x$. Distributions
with decreasing mortality rates have heavy tails. Distributions with increasing
mortality rates have light tails.

\subsection{Cited reference percentile: VaR and TVaR\label{subs}}
$\mbox{VaR}_p(X)$ is the value of $x_p$ such that {$\Pr(X>x_p)=p$, $0<p<1$,} i.e. the probability than the number of cites will exceed the VaR$_p$\%. Therefore, it is the 100$pth$ percentile of the random variable $X$.

{Value at Risk (VaR) has become the standard risk measure used to assess risk exposure in financial and actuarial issues. From the bibliometric point of view, we can make a simile considering the risk of the fact that a publication becomes obsolete. A researcher with high expectations regarding publications will prefer a disciplinary journal that becomes obsolete as late as possible. The survival function allows us to calculate the probability that the number of citations is greater than a specific value, say $x_p$. On the other hand, the VaR allows calculating the number of citations for which said probability is precise $p$. 

Suppose a decision maker, for example, a possible researcher interested in a certain disciplinary journal for publishing a paper. In that case, it should opt for the one with a fixed probability $p$ maximizes the value $x_p$. That is, between two disciplinary journals, say A and B, that produce the values at risk $x_p^A$ and $x_p^B$ for the exact value of $p$, respectively, the researcher should opt for the one with the highest value. This option will guarantee less obsolescence of the corresponding disciplinary journal. The choice of $p$ here is crucial and will depend on the consideration of each researcher.}

{ The measure VaR is merely a cutoff point and does not describe the tail behavior beyond the VaR threshold. The Tail-value-at-risk (TVaR) is a measure that is in many ways superior than VaR by reflecting the shape of the tail beyond VaR threshold.} The tail-value-at-risk at the 100$pth$ security
level, denoted by TVaR$_p(X)$, is the expected number of citations on the condition that the random variable $X$ exceeds the 100$pth$ percentile of $X$. That is,
{
\begin{eqnarray*}
\mbox{TVaR}_p(X)= E(X|X>x_p)=x_p+\frac{\sum_{x=x_p}^{\infty}(x-x_p)\Pr(X=x)}{R(x_p)}.
\end{eqnarray*}}

Some computations \citetext{see \citealp[Chap. 6]{klugmanetal_2008}} provide the TVaR for the NB distribution which is given by,
\begin{eqnarray*}
\mbox{TVaR}_p(X) = \frac{(x_p+1)\,_2F_1(1,x_p+1+\alpha,1+x_p,\beta/(\beta+1))}{
_2F_1(1,x_p+1+\alpha,2+x_p,\beta/(\beta+1))}.
\end{eqnarray*}

\section{Numerical experiments}
\begin{figure}[htbp]
\begin{center}
\includegraphics[page=17,width=0.65\linewidth]{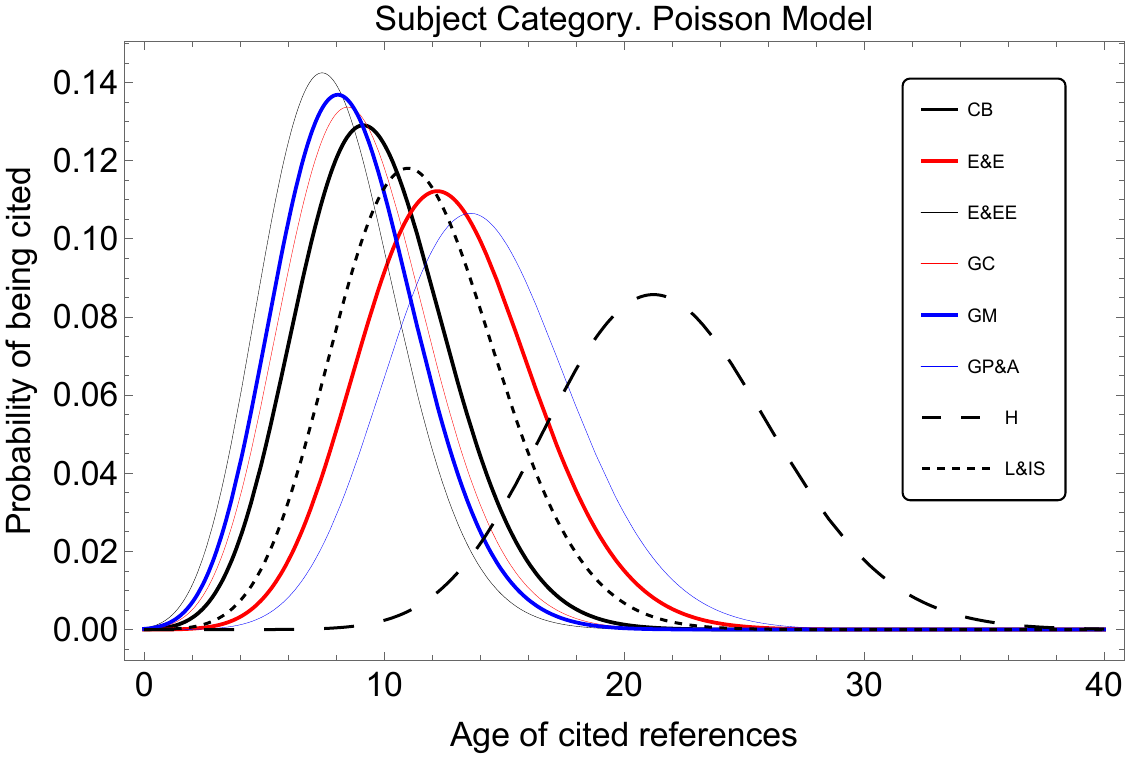}\\
\vspace*{1cm}
\includegraphics[page=18,width=0.65\linewidth]{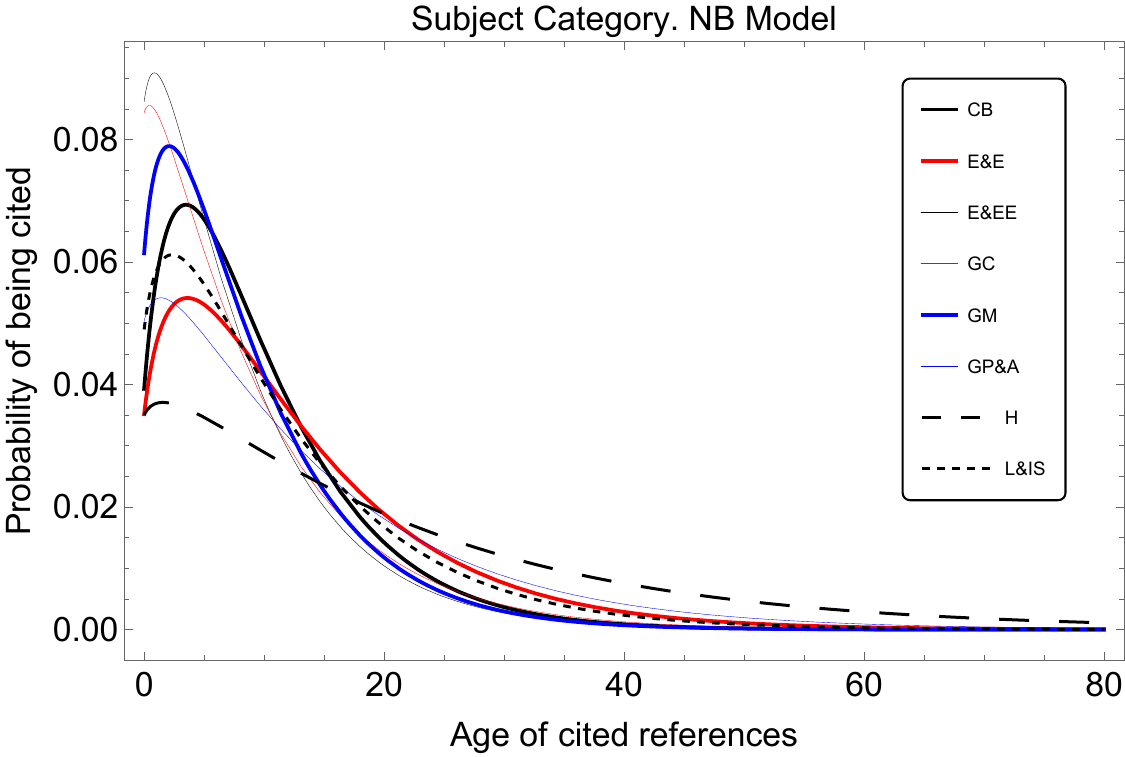}
\caption{{Fitted Poisson (above) and NB distribution (below)}\label{fig1}}
\end{center}
\end{figure}

Parameter estimates for the subject categories and the {disciplinary} journals by using the Poisson and negative binomial distributions are illustrated in Tables \ref{tab2} and \ref{tabA2} (Appendix A), respectively. We have used the Akaike's information criterion (AIC) as a measure of the model selection. A lower value of this measure
of model selection is desirable. {It is observable that the negative binomial distribution fits the data well, which is confirmed looking at the {Figure
\ref{fig1}.}}

\begin{figure}[htbp]
\begin{center}
\includegraphics[page=1,width=0.35\linewidth]{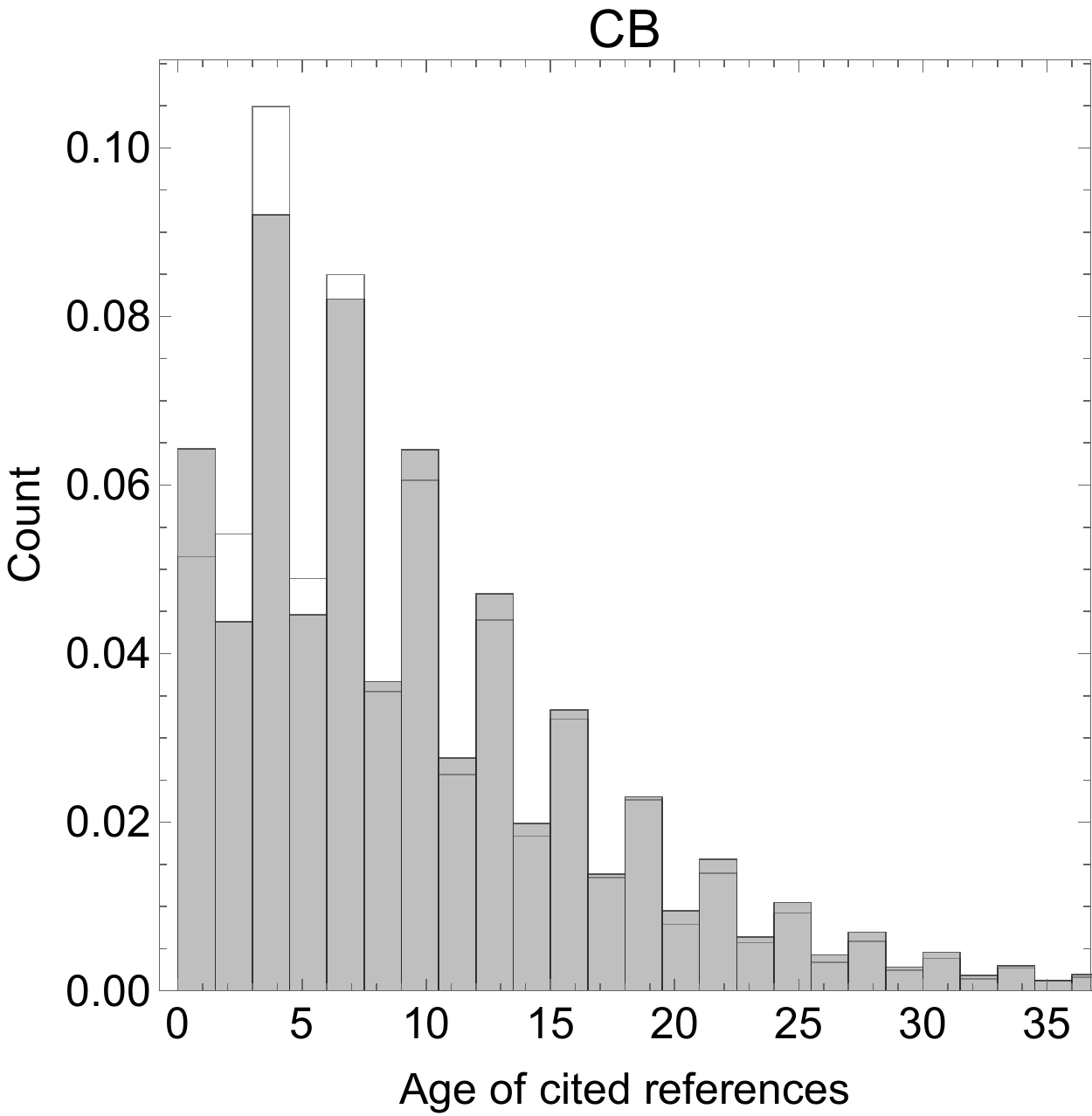}\hspace{0.25cm}\includegraphics[page=2,width=0.35\linewidth]{figures.pdf}\\
\includegraphics[page=3,width=0.35\linewidth]{figures.pdf}\hspace{0.25cm}\includegraphics[page=4,width=0.35\linewidth]{figures.pdf}\\
\includegraphics[page=5,width=0.35\linewidth]{figures.pdf}\hspace{0.25cm}\includegraphics[page=6,width=0.35\linewidth]{figures.pdf}\\
\hspace*{1.70cm}\includegraphics[page=7,width=0.35\linewidth]{figures.pdf}\hspace{0.25cm}
\includegraphics[page=8,width=0.44\linewidth]{figures.pdf}
\caption{Empirical and fitted distribution using NB model\label{fig2}}
\end{center}
\end{figure}

\begin{figure}[htbp]
\begin{center}
\includegraphics[page=9,width=0.35\linewidth]{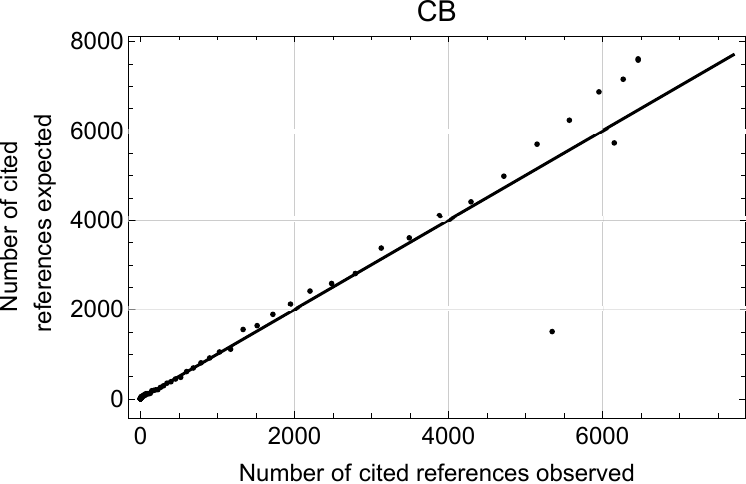}\hspace{0.25cm}\includegraphics[page=10,width=0.35\linewidth]{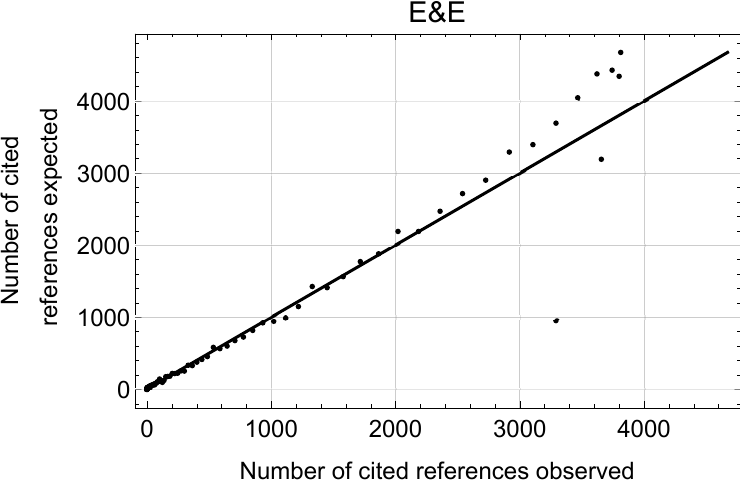}\\
\includegraphics[page=11,width=0.35\linewidth]{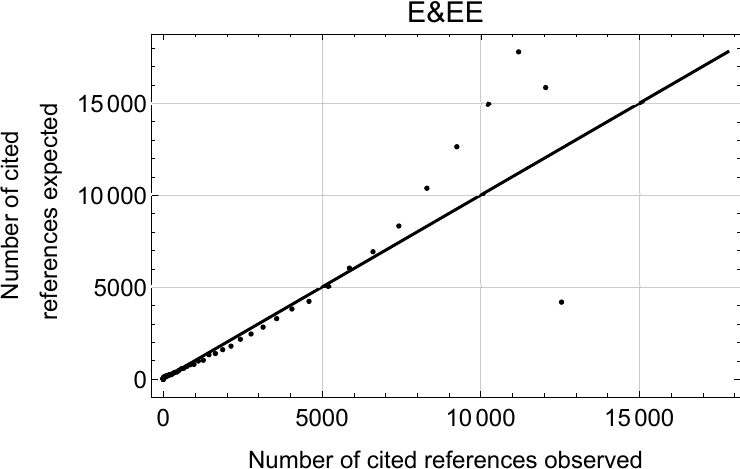}\hspace{0.25cm}\includegraphics[page=12,width=0.35\linewidth]{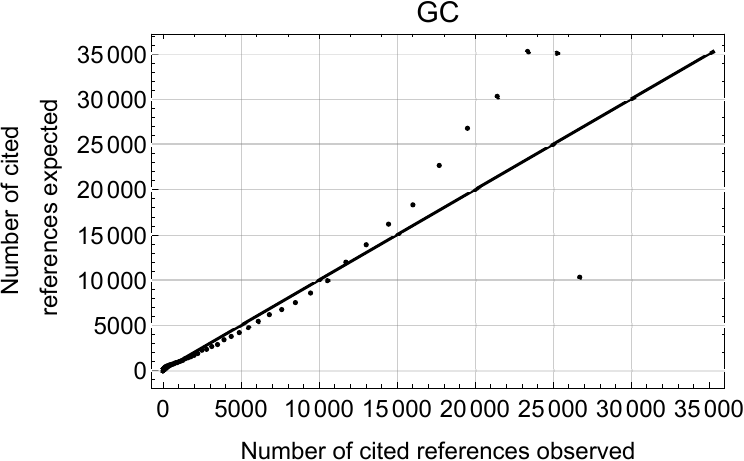}\\
\includegraphics[page=13,width=0.35\linewidth]{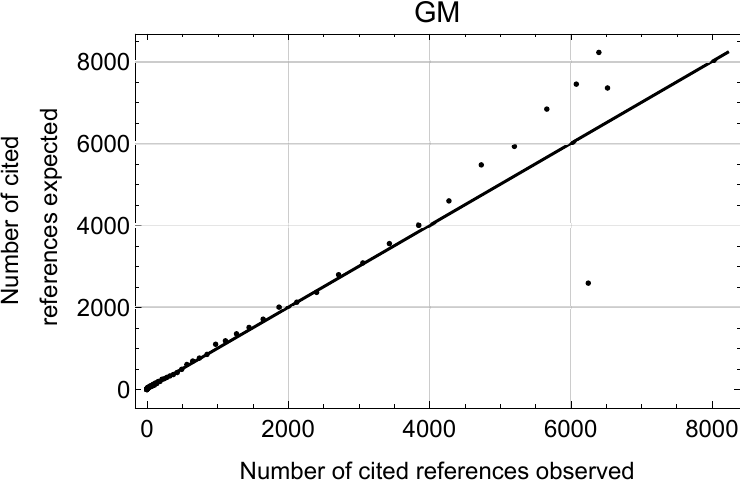}\hspace{0.25cm}\includegraphics[page=14,width=0.35\linewidth]{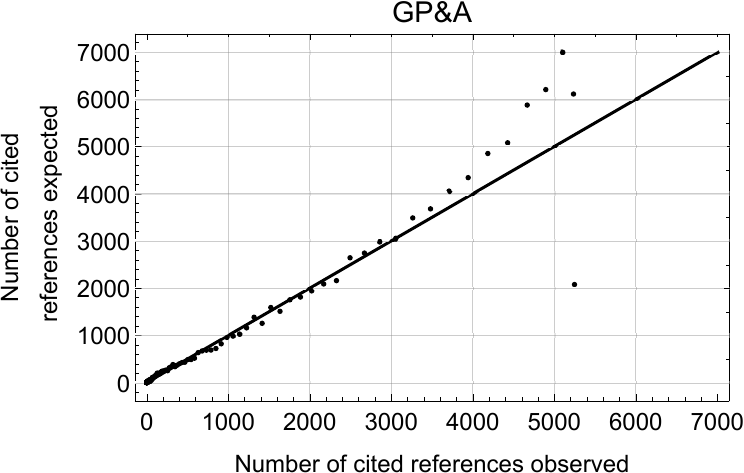}\\
\includegraphics[page=15,width=0.35\linewidth]{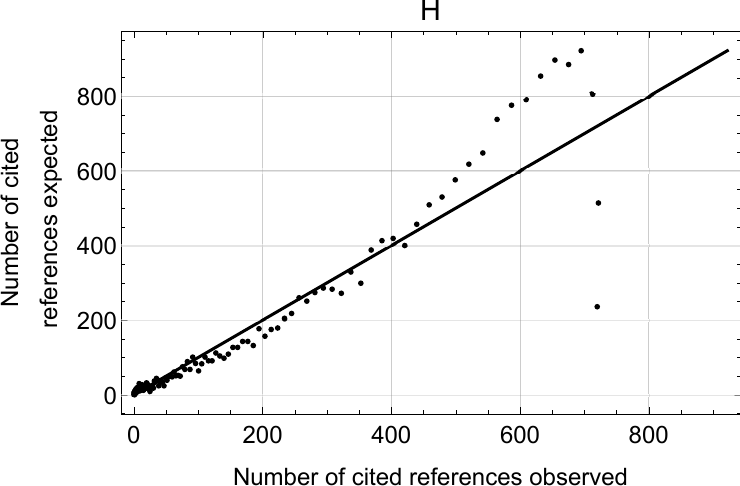}\hspace{0.25cm}\includegraphics[page=16,width=0.35\linewidth]{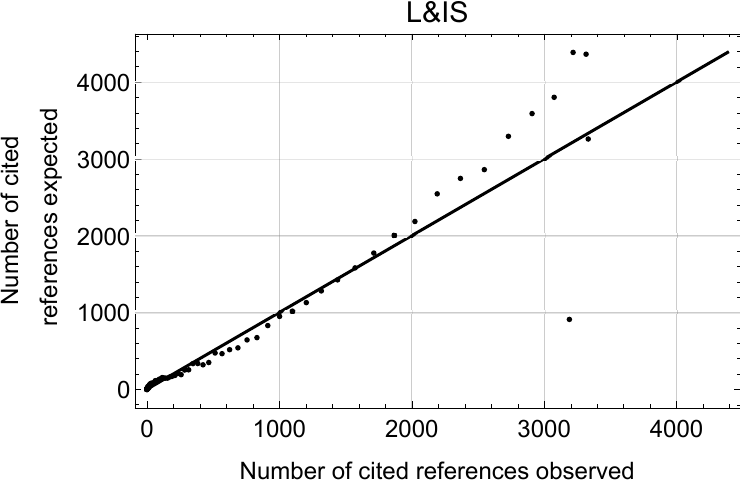}
\caption{Empirical ($-$) and fitted distributions using NB ($\cdot$) model\label{fig3}}
\end{center}
\end{figure}

The improvement in the fit of the NB model compared to the Poisson model is remarkable in all subject categories and journals analyzed. In the case of the subject categories (see Table \ref{tab2} in the Appendix), the reduction in the AIC for the NB model compared to the Poisson model is greater than 37\% for all categories, even reaching 67\% in the case of History. Furthermore, the variability in the parameters within each model reflects the differences observed both between the different subfields (Table \ref{tab2}) and between the journals within the same subfield (see Table \ref{tabA2} in the Appendix).

{The field effect can be seen better in {Figure \ref{fig1}}. The upper graph of this Figure shows the Poisson probability function and the negative binomial probability function in the lower part. For that, we have used the estimated parameters that appear in Table \ref{tab2}. It is observed that the latter better captures the pattern followed by the field that appears in {the empirical data}. Furthermore, the NB distribution approaches zero much more slowly, thus showing a much heavier right tail than the Poisson distribution. } Within each model, the differences between the distributions are considerable. According to the NB model, which offers a better fit to the empirical data, the differences are clearly observed both in the peak and in the tail of the distribution. The subfield with the heaviest tail, and therefore with the least obsolescence, is History. On the contrary, the subfields with the lightest tails, and therefore where the obsolescence is greater, correspond to Electrical \& Electronic Engineering, General Chemistry, General Medicine, and Cell Biology.

However, it is sometimes not obvious to compare the citation distribution for two different fields. This is because the curves often intersect each other on one or more occasions. It would therefore be convenient to use some measures for the tail of the citation distribution, which is the part of the curve that best defines the obsolescence of the literature in a field. {Since the accuracy obtained with the NB distribution is much better than that achieved with the Poisson distribution, from now on, we will limit ourselves to offering subsequent numerical results only for the first one.} { The accuracy obtained with the NB distribution is confirmed looking at the Figures \ref{fig2} and \ref{fig3}} The empirical application of the obsolescence measures proposed in this paper is shown in Tables \ref{tab3} to \ref{tab4} and Tables \ref{tabA3} to \ref{tabA5} (see Appendix).

\subsection{Cited reference survival rate}
As previously stated, the tail of the distribution allows to measure the obsolescence in a field. As can be seen in Table \ref{tab3} and Tables \ref{tabA3}-\ref{tabA4} in Appendix, there are very important differences between subfields in relation to the tail of the distribution. In the case of the subject categories (Table \ref{tab3}), the one with the highest survival rate in the cited references is History. Above 41\% of the cited references in the journals of History are over twenty years, and about 10\% are over fifty. {The second highest survival rate is achieved in General Physics \& Astronomy. Above 25\% of the cited references in the journals of General Physics \& Astronomy are over twenty years, and about 2.6\% are over fifty.}

The subject categories with the lowest survival rates are Electrical \& Electronic Engineering and General Medicine. Only 8\% of the cited references in Electrical \& Electronic Engineering are over twenty years, while 0.1\% are over fifty. In the case of General Medicine, only 9\% of the cited references in the journals of this category are over twenty years, and 0.1\% are over fifty.

Therefore, the obsolescence in Electrical \& Electronic Engineering is the order of five times higher than the obsolescence in History at medium time period (twenty years) but the order of eighty times higher at long time period (fifty years). Note that this proportion increases exponentially over time. Thus, for example, after a hundred years this proportion is 7000 to 1, that is, the survival rate for Electrical \& Electronic Engineering is approximately 7000 times lower than that of History.

Although two subfields very far in their citation habits (Electrical \& Electronic Engineering and History) have been compared, important differences can also be observed between subject categories of the same branch of knowledge. This is the case, for example, in science between General Physics \& Astronomy and General Chemistry. In General Physics \& Astronomy, over 25\% of the citations are directed to publications over twenty years. This percentage is reduced to 11\% in General Chemistry, which is less than half that in General Physics \& Astronomy. On the other hand, these proportions again increase exponentially over time. Thus, for example, after fifty years, the relationship is approximately 7 to 1, that is, the survival rate in General Chemistry is 7 times lower than in General Physics \& Astronomy. However, at a hundred years this proportion is approximately 50 to 1, with a survival rate for General Chemistry being around fifty times lower than for General Physics \& Astronomy.

These differences between subfields, although with lower proportions, can also be observed between other subfields and even between journals of the same subject category (see Tables \ref{tabA3} and \ref{tabA4} in the Appendix). Although the subject category is the smallest disaggregation set in relation to the field in the Scopus database, it is common for the same subject category to include very different specialties in their referencing habits. As an example, we can analyze the tail of the distribution for two different specialties within the Library \& Information Sciences category. The journal Scientometrics clearly shows less obsolescence than Journal of Information Science. While the survival rate after twenty years for the first journal is around 1.7 times that the second one, the survival rate after sixty years is around 22 times higher for Scientometrics, and around 300 times higher after a hundred years.

Something similar happens with journals in other subject categories. In General Medicine, for example, the obsolescence in New England Journal of Medicine is much lower than that of PLoS Medicine. While the survival rate after twenty years for the first journal is around 1.5 times that the second one, the survival rate after sixty years is around 5.5 times higher for the first journal, and around 32 times higher after a hundred years.

\subsection{Cited reference mortality rate}
As previously evidenced, the survival rate decreases rapidly as the age of the cited reference increases. However, the mortality rate (hazard rate) takes on more stable values. It can be observed in Table \ref{tab3} that all the analyzed subject categories present increasing mortality rates, which indicates that their distributions have light tails. The tails will be lighter the higher their mortality rates are. On the contrary, they will be less light the lower their mortality rates are. Thus, History has the lowest mortality rates (around 0.04). Although the mortality rate increases slightly over time, the range of variation is quite narrow (between 0.0455 and 0.0476). With a mortality rate that is also low (around 0.07) is General Physics \& Astronomy. However, the highest mortality rates (greater than 0.12) are reached in Cell Biology, General Medicine, and Electrical \& Electronic Engineering.

Again there are important differences between subject categories. For example, as extreme cases, the mortality rates in Electrical \& Electronic Engineering are about three times those in History. Within the same branch, the differences are minor, although they remain remarkable. Thus, the mortality rates in General Chemistry are approximately 50\% higher those in General Physics \& Astronomy.

If we again compare the journals Scientometrics and Journal of Information Science (Table \ref{tabA4} in the Appendix), the mortality rates of the cited references in the second journal are approximately 50\% higher than those in Scientometrics. In General Medicine, the mortality rates of the cited references in PLoS Medicine are between 25\% and 29\% higher than those in New England Journal of Medicine.

\subsection{Cited reference percentile: VaR and TVaR}
A lower obsolescence, this is a slower aging, is associated with higher values for a given percentile (see Tables \ref{tab4} and \ref{tabA5} in the Appendix). Of the subject categories analyzed, the one that presents a minor obsolescence attending to the percentiles, this is a slower aging of the literature, is History. Attending TVaR, a 5\% of the cited literature in History is more than 65 years, while 1\% is over 98. At a distance is General Physics \& Astronomy, where 5\% of the cited literature is over 42 years, and 1\% is over 63.

On the contrary, a higher obsolescence, that is a faster aging, is associated with lower values for a given percentile. Of the subject categories analyzed, those that present a major obsolescence, this is a fast aging of the literature, are Electrical \& Electronic Engineering, General Medicine, Cell Biology, and General Chemistry. Thus, a 5\% of the cited literature in Electrical \& Electronic Engineering is over 24 years, while 1\% is over 36.

At the level of disciplinary journals (Table \ref{tabA5} in the Appendix), in general the differences within the same subject category are less than between journals from different categories. As an example of two different disciplines within the same subject category, we can analyze the case of the Library \& Information Sciences. The journal Scientometrics clearly shows less obsolescence than Journal of Information Science. Again attending the TVaR, 5\% of the literature cited in Scientometrics is over 31 years, while 1\% is over 46. However, in the Journal of Information Science, 5\% of the literature is over 24 years, and 1\% is over 34.

{Let us now suppose that a researcher with a career in Economics decided to opt for one of the two journals, Applied Economics or Economics Letters, which are very similar in the content they disseminate (the second is more practical). Following the comments developed in subsection \ref{subs} and if you set a probability $p=0.01$, that is, that the distribution of the number of citations leaves an area to the right of 1\%, you should choose to send the article to the first journal, which has a VaR with a value of 60 instead of the VaR 49 that the latter has (see Table \ref{tabA5}). If the researcher considers that the value of $p$ that measures journal obsolescence should be 9\%, then there is not much difference between the two VaR values (32 and 29), and it could feel indifferent in this case.}

Furthermore, the following comparison can also be made. While in Scientometrics there are 9\% of cited references with more than 26 years, in Journal of Information Science with this same age the percentage is reduced to only 4\%. Similarly, while in Scientometrics there are 6\% of cited references with more than 30 years, in the Journal of Information Science with this same age the percentage of cited references is reduced to only 2\%.

In General Medicine, 5\% of the literature cited in New England Journal of Medicine is over 26 years, and 1\% is over 38. However, in PLoS Medicine, 5\% of the literature is over 22 years, and 1\% is over 32. Moreover, while in New England Journal of Medicine there are 9\% of cited references with more than 21 years, in PLoS Medicine with this same age the percentage is reduced to only 6\%.

In Economics \& Econometrics, 5\% of the literature cited in American Economic Review is over 43 years, and 1\% is over 65. However, in Applied Economics, 5\% of the literature is over 35 years, and 1\% is over 50. Furthermore, while in American Economic Review there are 6\% of cited references with more than 40 years, in Applied Economics with this same age the percentage is reduced to only 3\% of the cited references.

Therefore, a detailed analysis of Table \ref{tabA5} (see the Appendix) would allow, although it is not the purpose of this work, to group journals with similar obsolescence through the analysis of the tails of their distributions. Similarly, it would also allow subject categories to be disaggregated into different disciplines by analyzing obsolescence through the behavior of the tails in the distributions.

\begin{table}[htbp]
  \resizebox{0.8\textwidth}{!}{\begin{minipage}{\textwidth}
       \caption{{From top to down survival rate and mortality rate for the subject categories}\label{tab3}}
\begin{center}
\begin{tabular}{lccccccccc}\hline
& \multicolumn{9}{c}{Age}\\ \cline{2-10}
 & 20 & 30 & 40 & 50 & 60 & 70 & 80 & 90 & 100  \\
\cline{2-10}

CB & 0.1112 & 0.0273 & 0.0063 & 0.0014 & 0.0003 & 0.0001 & 1.4E-5    & 3.01E-6 & 6.30E-7 \\
   & 0.1278 & 0.1341 & 0.1377 & 0.1400 & 0.1416 & 0.1428 & 0.1437    & 0.1445   & 0.1451 \\

E\&E & 0.2115    & 0.0813    & 0.0302    & 0.0109    & 0.0039    & 0.0013  & 0.0004  & 0.0001 & 6.0E-5 \\
     & 0.0891    & 0.0931    & 0.0954    & 0.0969    & 0.0980    & 0.0988  & 0.0994  & 0.0999 & 0.1003 \\

E\&EE & 0.0826 & 0.0212 & 0.0053 & 0.0013 & 0.0003 & 8.2E-5 & 2.03E-5 & 5.00E-6  & 1.22E-6 \\
      & 0.1259 & 0.1279 & 0.1290 & 0.1297 & 0.1302 & 0.1305 & 0.1308  & 0.1310   & 0.1312 \\

GC & 0.1153 & 0.0368 & 0.0116 & 0.0036 & 0.0011 & 0.0003 & 0.0001 & 3.4E-5    & 1.07E-5 \\
   & 0.1073 & 0.1084 & 0.1091 & 0.1095 & 0.1097 & 0.1100 & 0.1101 & 0.1103    & 0.1104 \\

GM & 0.0910  & 0.0222 & 0.0052    & 0.0012 & 0.0002 & 6.2E-5 & 1.4E-5 & 3.13E-6 & 6.95E-7 \\
   & 0.1292  & 0.1332 & 0.13547  & 0.1369 & 0.1378 & 0.1386 & 0.1391 & 0.1396  & 0.1400 \\

GP\&A & 0.2539  & 0.1206  & 0.0566 & 0.0264  & 0.0123  & 0.0056  & 0.0026  & 0.0012  & 0.0005 \\
     & 0.0711  & 0.0723  & 0.0731 & 0.0735  & 0.0739  & 0.0741  & 0.0743  & 0.0745  & 0.0746 \\

H & 0.4151  & 0.2595  & 0.1612  & 0.0998  & 0.0616  & 0.0380  & 0.0233  & 0.0143  & 0.0088 \\
  & 0.0455  & 0.0462  & 0.0466  & 0.0469  & 0.0471  & 0.0473  & 0.0474  & 0.0475  & 0.0476 \\

L\&IS & 0.1793  & 0.0663  & 0.0239  & 0.0085  & 0.0030  & 0.0010 & 0.0003 & 0.0001 & 4.3E-5 \\
      & 0.0932  & 0.0960  & 0.0976  & 0.0987  & 0.0994  & 0.0999 & 0.1004 & 0.1007 & 0.1010 \\
\hline
 \end{tabular}
 \end{center}
 \end{minipage}}
 \end{table}

\begin{table}[htbp]
\caption{Percentile VaR (above) and TVaR (below) for subject categories\label{tab4}}
\begin{center}
\begin{tabular}{lccccccccc}\hline
& \multicolumn{9}{c}{$p$}\\ \cline{2-10}
 & 0.01 & 0.02 & 0.03 & 0.04 & 0.05 & 0.06 & 0.07 & 0.08 & 0.09 \\ \cline{2-10}
CB  & 36 & 32 & 29 & 27 & 25 & 24 & 23 & 22 & 21 \\
 & 37.18 & 33.18 & 30.18 & 28.18 & 26.18 & 25.18 & 24.18 & 23.18 & 22.18 \\
E\&E  &  50 & 44 & 40 & 37 & 34 & 33 & 31 & 30 & 28 \\
 & 51.11 & 45.11 & 41.11 & 38.11 & 35.11 & 34.11 & 32.11 & 31.11 & 29.11 \\
E\&EE   &  35 & 30 & 27 & 25 & 23 & 22 & 21 & 20 & 19 \\
 & 36.15 & 31.15 & 28.15 & 26.15 & 24.15 & 23.15 & 22.15 & 21.15 & 20.15 \\
GC &  41 & 35 & 31 & 29 & 27 & 25 & 24 & 23 & 22 \\
 & 42.12 & 36.12 & 32.12 & 30.12 & 28.12 & 26.12 & 25.12 & 24.12 & 23.12 \\
GM  &  35 & 30 & 27 & 25 & 24 & 22 & 21 & 20 & 20 \\
 & 36.16 & 31.17 & 28.17 & 26.17 & 25.17 & 23.17 & 22.17 & 21.17 & 21.17 \\
GP\&A   &  62 & 53 & 48 & 44 & 41 & 39 & 37 & 35 & 33 \\
 & 63.08 & 54.08 & 49.08 & 45.08 & 42.08 & 40.08 & 38.08 & 36.08 & 34.08 \\
H   &  97 & 83 & 74 & 68 & 64 & 60 & 57 & 54 & 52 \\
 & 98.05 & 84.05 & 75.05 & 69.05 & 65.05 & 61.05 & 58.05 & 55.05 & 53.05 \\
L\&IS &  48 & 41 & 37 & 34 & 32 & 30 & 29 & 28 & 26 \\
 & 49.11 & 42.11 & 38.11 & 35.11 & 33.11 & 31.11 & 30.11 & 29.11 & 27.11 \\ \hline

 \end{tabular}
 \end{center}
 \end{table}

\section{Conclusions}

The literature framework establishes that a document is obsolete when it is no longer cited, i.e., when it is no longer used by an academic community as a source of information to argue, justify or contradict the statements or findings reported by other authors.

The results in this study support that the subfield and even the discipline (specialty) are influencing obsolescence. Thus, there is a field effect in the phenomenon by which publications are less and less cited over time, known as literature obsolescence.

We used all the cited references in 22,559 research articles published in 2019 from eight different subfields (subject categories in Scopus): Cell Biology, Economics \& Econometrics, Electrical \& Electronic Engineering, General Chemistry, General Medicine, General Physics \& Astronomy, History, and Library \& Information Sciences.

The distribution of synchronically accumulated citations produced an initial growth followed by an exponential decrease. We concluded that the negative binomial is preferable to the Poisson distribution for the datasets considered in all the cases.

However, it is not obvious to compare the citation distribution for two different disciplinary journals. This is because the curves often intersect each other on one or more occasions. For this reason, {three measures to analyze the tail of the distribution were proposed: survival rate, mortality rate and percentile.}

There are very important differences between subfields and even between disciplines in relation to the tail of the distribution. The highest survival rate is observed in History. Above 41\% of the cited references are over twenty years, and about 10\% are over fifty. At a certain distance from this last subfield, although with a similarly high survival rate, is the General Physics \& Astronomy field, for which more than 25\% of the cited references have a survival rate of more than twenty years and close to 2.6\% over fifty. On the contrary, the lowest survival rates are observed in Electrical \& Electronic Engineering and General Medicine. Only 8\% of the cited references in Electrical \& Electronic Engineering and 9\% in General Medicine are over twenty years, while 0.1\% are over fifty. Therefore, the obsolescence in Electrical \& Electronic Engineering is the order of five times higher than in History at medium time period (twenty years) but the order of eighty times higher at long time period (fifty years). Note that this proportion increases exponentially over time.

Important differences are also observed between subfields of the same branch of knowledge. This is the case of General Physics \& Astronomy and General Chemistry, for example. A 25\% of the citations in General Physics \& Astronomy are directed to publications over twenty years, and this percentage is reduced to 11\% in General Chemistry (less than half). Furthermore, this proportion again increase exponentially over time. Thus, after fifty (a hundred) years, the survival rate in General Chemistry is seven (50) times lower than in General Physics \& Astronomy.

Differences are also observed at journal level. While the survival rate at twenty years in the journal Scientometrics is around 1.7 times that in Journal of Information Science, after sixty years it is around 22 times higher for Scientometrics. Something similar happens with journals in other subfields.

The mortality rate takes on more stable values than the survival rate. All the subfields present increasing mortality rates, which indicates that their distributions have light tails. History has the lowest mortality rates (around 0.04) followed by General Physics \& Astronomy (around 0.07). The highest mortality rates (greater than 0.12) are reached in Cell Biology, General Medicine, and Electrical \& Electronic Engineering. Again, there are important differences both between subfields and between disciplines. The mortality rates in Electrical \& Electronic Engineering are about three times those in History, and in General Chemistry are about 50\% higher than in General Physics \& Astronomy. At disciplinary journal level, the mortality rates in Journal of Information Science are about 50\% higher than those in Scientometrics, for example.

Actually, this is an expected result if we consider the diffusion process of knowledge in the different fields.  {Two main types of obsolescence occur related to the diffusion process. Both types are related to the rate at which diffusion occurs. In the first place, initially, there is a high number of citations until a modal value is reached, followed by a rapid drop in them, a drop that is also very sharp, as occurs, for example, in the fields of Medicine and Engineering. The second type is due to a slower diffusion process, associated with a slower rate of decline. Examples of this behavior in obsolescence are the basic sciences, social sciences, and humanities.}

Finally, a higher obsolescence, that is a faster aging, is associated with lower values for a given percentile. The highest obsolescence is again observed in Electrical \& Electronic Engineering, General Medicine, Cell Biology, and General Chemistry. Thus, attending the TVaR, a 5\% of the cited literature in Electrical \& Electronic Engineering is over 24 years, while 1\% is over 36. On the contrary, a 5\% of the cited literature in History is more than 65 years, while 1\% is over 98. At disciplinary journal level, while in Scientometrics there are 6\% of cited references with more than 30 years, in Journal of Information Science with this same age the percentage of cited references is reduced to only 2\%, for example.

As has been evidenced, the difference between subfields can also be observed between disciplinary journals of the same subfield. Although the subject category is the smallest disaggregation set in relation to the field in the Scopus database, it is common for the same subject category to include very different disciplines. The differences can be very noticeable in the use that an academic community makes of bibliographical references to argue, justify or contradict the statements or findings reported by other authors.

In this respect, as practical application and future line of research, a detailed analysis of the measures proposed in this paper would allow to group journals with similar obsolescence through the comparison of the tail in the distributions. Similarly, it would also allow subject categories to be disaggregated into different disciplines by analyzing obsolescence through the tail of the distributions.

The journal impact factor focus on measuring the average citations per document in a short period of time (between two and five years generally). These measures do not collect the citations received in long time periods. However, as has been evidenced in this paper, some journals with low obsolescence accumulate a high percentage of citations after many years. Therefore, it would be necessary to accompany the short-term impact factor with some measure for the tail of the citation distribution, such as those presented in this paper, to provide a more accurate idea of the real impact of said journal.

As a final consideration, diachronic analysis is much less common in the literature as it takes time for citation to accumulate. However, our approach can lead also to diachronic analysis because it could be applied in the same way. Diachronic analysis is based on citations received, instead of synchronic analysis which is based on references made. Therefore, diachronic obsolescence requires setting a period of time in the past to look into the future. In our approach but for the diachronic analysis, the years of the citations must be considered in natural chronological order, instead of the reverse chronological order of the synchronic analysis. However, some authors argue that synchronic and diachronic studies produce similar results \citetext{\citealp{Stinson1987}}, suggesting a preference for the synchronic method.

\section*{Appendix}

\begin{sidewaystable}
  \resizebox{0.5\textwidth}{!}{\begin{minipage}{\textwidth}
       \caption{Journals and subject categories in the empirical application\label{tabA1}}
\begin{center}
\begin{tabular}{llll}
Abbreviated journal title & Full journal title & Abbreviated journal title & Full journal title\\ \hline
Cell Biology (CB) &   & Economics and Econometrics (E\&E) &\\ \cline{1-1} \cline{3-3}
Ag.Cell              &       Aging Cell                       &  Amer. Econ. Rev.       &   American Economic Review \\
Autoph.              &       Autophagy                        &  Appl. Econ.            &   Applied Economics         \\
Blood                &       Blood                            &  Appl. Econ. Letters    &   Applied Economics Letters \\
Canc. C.             &       Cancer Cell                      &  Ecol. Econ.            &   Ecological Economics    \\
C. Metab.            &       Cell Metabolism                  &  Econ. Jour.            &   Economic Journal        \\
Jour. C. Biol.       &       Journal of Cell Biology          &  Econ. Letters          &   Economics Letters        \\
Nat. Cell Biol.      &       Nature Cell Biology              &  Jour. of Econ. Persp.  &   Journal of Economic Perspectives   \\
Nat. Meth.           &       Nature Methods                   &  Jour. of Finan. Econ.  &   Journal of Financial Economics     \\
Plant C.             &       Plant Cell                       &  Quart. Jour. Econ.     &   Quarterly Journal of Economics      \\
Plant Jour.          &       Plant Journal                    &  Small Bus. Econ.       &   Small Business Economics   \\ \\ \hline

Electrical and Electronic Engineering (E\&EE) & &   General Chemistry (GC) &\\ \cline{1-1} \cline{3-3}
IEEE Comm. Mag.      &        IEEE Communications Magazine    &   Acc. Chem. Res.       &  Accounts of Chemical Research   \\
IEEE Sens. Jour.     &        IEEE Sensors Journal            &   ACS Cat.              &  ACS Catalysis                   \\
IEEE Trans. Cyb.     &        IEEE Transactions on Cyberneti  &   Adv. Fun. Mat.        &  Advanced Functional Materials    \\
IEEE Trans. VT       &        IEEE Transactions on VT         &   Chem                  &  Chem                              \\
IEEE Wir. Comm.      &        IEEE Wireless Communications    &   Chem. Rec.            &  Chemical Record                   \\
Nano En.             &        Nano Energy                     &   Chem. Sc.             &  Chemical Science                   \\
Nat. Electr.         &        Nature Electronics              &   Chem. Mat.            &  Chemistry of Materials             \\
Nat. Nanot.          &        Nature Nanotechnology           &   Jour. Mat. Chem. B    &  Journal of Materials Chemistry B   \\
Procc. IEEE          &        Proceedings of the IEEE         &   Nano Let.             &  Nano Letters                        \\
Sig. Process.        &        Signal Processing               &   Nat. Mat.             &  Nature Materials                 \\ \\ \hline

General Medicine (GM) & & General Physics and Astronomy (GP\&A) &\\ \cline{1-1} \cline{3-3}
BMC Med.             &   BMC Medicine                                       & Adv. in Theor. \& MP     &  Advances in Theoretical and Mathematical Physics        \\
Clin. Sc.            &   Clinical Science                                   & Chaos                    &  Chaos                                                   \\
Front.  Med.         &   Frontiers in Medicine                              & Comb. and Flame          &  Combustion and Flame                                      \\
JAMA-Jour. AMA       &   JAMA Journal of the American Medical Association   & Comm. in Number T\&P     &  Communications in Number Theory and Physics               \\
Jour.  Clin. Inv.    &   Journal of Clinical Investigation                  & Comp. Physics Commun.    &  Computer Physics Communications                           \\
Jour.  Exp. Med.     &   Journal of Experimental Medicine                   & Jour. Biophot.        &  Journal of Biophotonics                                    \\
New Eng. Jour. Med.  &   New England Journal of Medicine                    & Jour. Comp. Physics   &  Journal of Computational Physics                           \\
PLoS Medicine        &   PLoS Medicine                                      & Jour. Physical \& CRD &  Journal of Physical and Chemical Reference Data            \\
The BMJ              &   The BMJ                                            & Jour. Phys. A: M\&T   &  Journal of Physics A: Mathematical and Theoretical          \\
The Lancet           &   The Lancet                                         & Nat. Phys.               &  Nature Physics                                           \\ \\   \hline

History (H) &   & Library and Information Sciences (L\&IS) &\\ \cline{1-1} \cline{3-3}

Am. Ant.                           &  American Antiquity                                             &  IEEE Trans. Inf. Th.       &  IEEE Transactions on Information Theory                                \\
As. Stud. Rev.                     &  Asian Studies Review                                           &  Inf. Syst. Manag.          &   Information Systems Management                                        \\
Hist. Antrop.                      &  History and Antropology                                        &  Inf. Syst. Res.            &   Information Systems Research                                           \\
Hist. the Human Sc.             &  History of the Human Sciences \& Jour. of Infor. Sc.              &  Int. Jour. Inf. Manag.     &   International Journal of Information Management                         \\
Jour. East. Afr. Stud.          &  Journal of Eastern African Studies \& Jour. of Inform.            &  Jour. Inf. Sc.             &   Journal of Information Science                                          \\
Jour. Philos. Educ.          &  Journal of Philosophy of Education \& Jour. of the Ass. for ICT      &  Jour. Inform.              &   Journal of Informetrics                                                  \\
Latin Amer. Antiq.                 &  Latin American Antiquity                                       &  Jour. Ass. Inf. Sc. Tech.  &   Journal of the Association for Information Science and Technology        \\
Past. Pres.                        &  Past and Present                                               &  Soc. Sc. Comp. Rev.        &   Social Science Computer Review                                           \\
Philos. Sc.                     &  Philosophy of Science \& Social Sc. Comp. Rev.                    &  Scient.                    &   Scientometrics                                                            \\
Stud. H\&O Sc. A         &  Studies in History and Philosophy of Science Part A                      &  Res. Eval.                 &   Research Evaluation                                                    \\
\hline
\end{tabular}
\end{center}
\end{minipage}}
\end{sidewaystable}

\begin{table}[htbp]
\caption{Estimates of the parameters and AIC values for the different subject categories studied\label{tab2}}
\begin{center}
\begin{tabular}{lrrrrrr}\hline
 & \multicolumn{2}{c}{Poisson} && \multicolumn{3}{c}{NB}\\ \cline{2-3}\cline{5-7}
 & $\widehat\theta$ & AIC && $\widehat\alpha$ & $\widehat\beta$ & AIC\\ \cline{2-3}\cline{5-7}
CB      & 9.63  & 967474  && 1.71 &  0.18 & 611010\\
E\&E    & 12.72 & 955040  && 1.48 &  0.12 & 500757\\
E\&EE   & 7.92  & 1565420 && 1.21 &  0.15 & 863077\\
GC      & 8.98  & 4003250 && 1.13 &  0.12 & 2052140\\
GM      & 8.57  & 860744  && 1.44 &  0.17 & 525562\\
GP\&A   & 14.10 & 1610500 && 1.16 &  0.08 & 714828\\
H       & 21.73 & 484209  && 1.11 &  0.05 & 159815\\
L\&IS   & 11.50 & 741742  && 1.33 &  0.11 & 378400\\ \hline
\end{tabular}
\end{center}
\end{table}

\begin{table}[htbp]
  \resizebox{0.5\textwidth}{!}{\begin{minipage}{\textwidth}
       \caption{Estimates of the parameters and AIC values for the different journals between subject categories studied\label{tabA2}}

\begin{center}
\begin{tabular}{lrrrrrrlrrrrrrr}\hline
Journal & \multicolumn{2}{c}{Poisson} && \multicolumn{3}{c}{NB} & Journal & \multicolumn{2}{c}{Poisson} && \multicolumn{3}{c}{NB}\\ \cline{2-3}\cline{5-7} \cline{9-10}\cline{12-14}
 & $\widehat\theta$ & AIC && $\widehat\alpha$ & $\widehat\beta$ & AIC & & $\widehat\theta$ & AIC && $\widehat\alpha$ & $\widehat\beta$ & AIC\\ \cline{2-3}\cline{5-7} \cline{9-10}\cline{12-14}
Ag.Cell           &  9.21 & 61920.80 &&   1.99 & 0.21 & 42425.50  &  Amer. Econ. Rev.       & 13.68 & 71880.30 & & 1.03& 0.07 & 29571.10 \\
Autoph.           &  8.98 & 46577.70 &&   2.54 & 0.28 & 34284.70  &  Appl. Econ.            & 13.43 & 206165.00 & & 1.76& 0.13 & 117052.00 \\
Blood             &  9.27 & 165245.0 &&   1.51 & 0.16 & 101291.00 &  Appl. Econ. Letters    & 12.97 & 57740.00  & & 1.51& 0.11 & 30166.40 \\
Canc. C.          &  7.52 & 41931.30 &&   1.94 & 0.25 & 29639.40  &  Ecol. Econ.            & 10.72 & 217689.00 & & 1.49& 0.13 & 119008.00 \\
C. Metab.         &  9.26 & 94099.40 &&   1.59 & 0.17 & 56428.90  &  Econ. Jour.            & 15.06 & 62753.50 & & 1.74& 0.11 & 32780.20 \\
Jour. C. Biol.    & 10.82 & 134508.0 &&   1.88 & 0.17 & 86890.30  &  Econ. Letters          & 13.12 & 75747.40 & & 1.09& 0.08 & 34518.80 \\
Nat. Cell Biol.   &  8.55 & 56299.30 &&   1.72 & 0.20 & 37248.00  &  Jour. of Econ. Persp.  & 13.25 & 25301.90 & & 0.78& 0.05 & 7863.46 \\
Nat. Meth.        &  7.56 & 57541.90 &&   1.23 & 0.16 & 33402.10  &  Jour. of Finan. Econ.  & 13.30 & 71736.90 & & 1.59& 0.12 & 39294.60 \\
Plant C.          & 10.53 & 98847.80 &&   1.88 & 0.17 & 63521.20  &  Quart. Jour. Econ.     & 12.92 & 25276.80 & & 1.11& 0.08 & 11317.70 \\
Plant Jour.       & 10.67 & 199413.00 &&   1.78 & 0.16 & 123403.00 & Small Bus. Econ.       & 12.89 & 131918.00 & & 1.77& 0.13 & 77058.60 \\
\\

IEEE Comm. Mag.     &  4.68 & 21884.30   & &   1.27 & 0.27 & 13981.00              &   Acc. Chem. Res.     &  8.59 & 219046.00  & & 1.08 & 0.12  & 112400.00 \\
IEEE Sens. Jour.          &  8.75 & 444603.00   & &   1.33 & 0.15 & 253091.00      &   ACS Cat.            &  10.08 & 704621.00 & & 1.13 & 0.11  & 355198.00 \\
IEEE Trans. Cyb.   &  9.13 & 142672.00   & &   1.72 & 0.18 & 87760.40              &   Adv. Fun. Mat.      &  7.12 & 567205.00  & & 1.17 & 0.16  & 319208.00 \\
IEEE Trans. VT                &  7.09 & 260898.00   & &   1.27 & 0.17 & 154303.00  &   Chem                &  9.99 & 95778.30  & & 1.03 & 0.10  & 42998.80 \\
IEEE Wir. Comm.               &  4.06 & 11839.00    & &   1.58 & 0.38 & 8648.93    &   Chem. Rec.          &  9.92 & 48032.70  & & 1.58  & 0.16 & 28315.90 \\
Nano En.                      &  6.12 & 319068.00   & &   1.11 & 0.18 & 180322.00  &   Chem. Sc.           &  10.22 & 828813.00 & &  1.12 & 0.11 & 405662.00 \\
Nat. Electr.            &  7.78 & 24525.40   & &   1.21 & 0.15 & 13783.10          &   Chem. Mat.          &  9.31 & 608256.00  & & 1.03 & 0.11 & 290984.00 \\
Nat. Nanot.                   &  8.49 & 47967.30   & &   1.24 & 0.14 & 26290.00    &   Jour. Mat. Chem. B  &  7.35 & 285010.00  & & 1.37 & 0.18 & 176781.00 \\
Procc. IEEE                   & 10.28  & 125421.00  & &   0.90 & 0.08& 48508.20    &   Nano Let.           &  8.77 & 525844.00  & & 1.16 &  0.13 & 277006.00 \\
Sig. Process.                 & 10.22 & 125464.00   & &   1.40 & 0.13& 69401.70    &   Nat. Mat.           &  9.37 & 70806.30  & & 1.25 & 0.13  & 37005.00\\
\\
BMC Med.                                     &  8.22 & 83089.30 & &  1.45 & 0.17 & 51546.40      & Adv. in Theor. \& MP    &  17.58 & 28589.8 & & 1.34 & 0.07  & 12752.40 \\
Clin. Sc.                                 &  8.44 & 52946.50 & &  1.63 & 0.19 & 34700.9          & Chaos                                               &  12.91 & 249949. & & 1.08 & 0.08  & 112862.00 \\
Front.  Med.         &  9.45 & 55111.20 & &  1.28 & 0.13 & 30989.50                              & Comb. and Flame                                &  13.74 & 242256.00 & & 1.13 & 0.08  & 109938.00 \\
JAMA-Jour. AMA         &  8.34 & 51128.80 & &  1.29 & 0.15 & 30059.40                            & Comm. in Number T\&P        &  17.13 & 11940.  & & 1.09 &  0.06 & 4846.52 \\
Jour.  Clin. Inv.    &  9.35 & 192079.00 & &  1.79 & 0.19 & 125070.00                            & Comp. Physics Commun.                     &  14.90 & 174309.00 & & 1.29 & 0.08  & 80013.00 \\
Jour.  Exp. Med.     &  9.32 & 91381.10 & &  1.73 & 0.18 & 58490.80                              & Jour. Biophot.                                &  9.941 & 110198.00 & & 1.54 &  0.15 & 63478.40 \\
New Eng. Jour. Med.    &  8.32 & 93445.50 & &  1.16 & 0.14 & 51970.60                            & Jour. Comp. Physics                       &  14.21 & 357495.00 & & 1.26 &  0.08 & 165764.00 \\
PLoS Medicine                                    &  7.54 & 89142.10 & &  1.43 & 0.19 & 56684.00  & Jour. Physical \& CRD        &  27.56 & 16742.70 & & 1.15 &  0.04 & 5891.92 \\
The BMJ                                          &  7.87 & 84127.70 & &  1.15 & 0.14 & 48457.30  & Jour. Phys. A: M\&T      &  16.94 & 314333.00 & & 1.06 &  0.06 & 124159.00 \\
The Lancet                                       &  7.81 & 63725.20 & &  1.20 & 0.15 & 36063.00  & Nat. Phys.                                      &  11.89 & 70481.20 & & 1.11 &  0.09 & 32008.90 \\
\\
Am. Ant.                           & 21.67  & 40815.70 & &  1.15 &  0.05 & 14954.80         & IEEE Trans. Inf. Th.                        & 14.64 & 172890.00 & & 1.14 &  0.07 & 72734.40 \\
As. Stud. Rev.                         & 14.91  & 34926.30 & &  1.34 &  0.09 & 15502.60     & Inf. Syst. Manag.                                   & 12.22 & 16811.10 & & 1.95  &  0.16  & 10230.30 \\
Hist. Antrop.                        & 23.76  & 74794.00  & &  1.00 &  0.04 & 21854.50      & Inf. Syst. Res.                                       & 13.93 & 54318.90 & & 1.90 &  0.13  & 31520.10 \\
Hist. the Human Sc.                 & 31.23  & 66806.70 & &  1.03 &  0.03 & 17583.40        & Int. Jour. Inf. Manag.             &  9.84 & 128778.00 & &  1.40&  0.14  & 74777.60 \\
Jour. East. Afr. Stud.            & 12.90  & 18574.60 & &  1.54 &  0.12 & 10032.20          & Jour. Inf. Sc.                                         &  9.04 & 30916.30 & &  1.84&  0.20  & 19722.40 \\
Jour. Philos. Educ.               & 21.96  & 39162.10 & &  1.11 &  0.05 & 13366.40          & Jour. Inform.                                              & 11.23 & 40188.80 & & 1.25 &  0.11  & 19858.10 \\
Latin Amer. Antiq.                   & 21.59  & 37674.30 & &  1.32 &  0.06 & 14656.00       & Jour. Ass. Inf. Sc. Tech.                                 & 12.09 & 56047.80 & &  1.75&  0.14  & 31780.80 \\
Past. Pres.                                & 24.38  & 42560.20 & &  1.21 &  0.04 & 13743.90 & Soc. Sc. Comp. Rev.                                     & 10.62 & 14952.30 & & 1.56 &  0.14  & 8689.10 \\
Philos. Sc.                         & 19.15  & 52320.70 & &  1.12 &  0.05 & 18445.10        & Scient.                                                      & 10.45 & 152399.00 & & 1.22 &  0.11  & 75890.80 \\
Stud. H\&O Sc. A       & 23.60  & 56336.10 & &  1.13 &  0.04 & 18529.30                     & Res. Eval.                                               & 10.30 & 58012.30 & & 1.31 &  0.12   & 30804.50 \\ \hline

\end{tabular}
\end{center}
\end{minipage}}
\end{table}

{\tiny
\begin{landscape}
\begin{center}
\begin{longtable}{lrrrrrclrrrrr}

 \caption{{From top to down survival rate and mortality rate for the journals considered in CB, E\&E, E\&EE, and GC}\label{tabA3}}

\\
\hline
& \multicolumn{5}{c}{$Age$}  &&& \multicolumn{5}{c}{$Age$} \\ \cline{2-6}\cline{9-13}
& 20 & 40 & 60 & 80 & 100  &&& 20 & 40 & 60 & 80 & 100\\ \cline{2-6}\cline{9-13}

\hline
\endfirsthead
\multicolumn{12}{c}%
{\tablename\ \thetable\ -- \textit{Continued from previous page}} \\
\hline
& \multicolumn{5}{c}{$Age$}  &&& \multicolumn{5}{c}{$Age$} \\ \cline{2-6}\cline{9-13}
& 20 & 40 & 60 & 80 & 100  &&& 20 & 40 & 60 & 80 & 100\\ \cline{2-6}\cline{9-13}

\hline
\endhead
\hline \multicolumn{13}{r}{\textit{Continued on next page}} \\
\endfoot
\hline
\endlastfoot

Ag.Cell &  0.0899 & 0.0031 & 8.9E-5 & 2.31E-6 & 5.6E-8 &&  Amer. Econ. Rev.  & 0.2440 & 0.0579 & 0.0136 & 0.0032 & 0.0007   \\
           &  0.1461 & 0.1601 & 0.1656 & 0.1685 & 0.1704 && & 0.0691 & 0.0695 & 0.0697 & 0.0698 & 0.0698 \\ \\

Autoph.  &  0.0693 & 0.0011 & 1.4E-5 & 1.43E-7 & 1.34-9 &&  Appl. Econ. & 0.2279 & 0.0292 & 0.0032 & 0.0003 & 3.3E-5 \\
           &  0.1724 & 0.1940 & 0.2023 & 0.2067 & 0.2094 && & 0.0920 & 0.10196 & 0.10604 & 0.1082 & 0.1096 \\  \\

Blood      & 0.1078 & 0.0070 & 0.0004 & 2.3E-57 & 1.25E-6 && Appl. Econ. Letters & 0.2183 & 0.0316 & 0.0041 & 0.0005 & 6.4E-5  \\
           & 0.1235 & 0.1307 & 0.1335 & 0.1350 & 0.1360 &&&  0.0882 & 0.0949 & 0.0976 & 0.0991 & 0.1001  \\  \\

Canc. C.  & 0.0479 & 0.0008 & 1.20E-5 & 1.56E-7 & 1.91E-9 && Ecolog. Econ.  & 0.1513 & 0.0146 & 0.0012 & 0.0001 & 8.9E-6    \\ 
             & 0.1751 & 0.1886 & 0.1938 & 0.1965 & 0.1982 && & 0.1062 & 0.1129 & 0.1156 & 0.1171 & 0.1180 \\  \\

C. Metab.  & 0.1046 & 0.0061 & 0.0003 & 1.6E-5 & 7.48E- && Econom. Jour.   & 0.2778 & 0.0458 & 0.0065 & 0.0008 & 0.0001     \\
                 & 0.1273 & 0.1356 & 0.1389 & 0.1406 & 0.1417 &&  & 0.0808 & 0.0903 & 0.0942& 0.0964 & 0.0977    \\    \\

Jour. C. Biol.  & 0.1420 & 0.0093 & 0.0005 & 2.6E-5  & 1.2E-6 && Econ. Letters   & 0.2292 & 0.0485 & 0.0101 & 0.0020 & 0.0004 \\
                         & 0.1200 & 0.1322 & 0.1371 & 0.1397  & 0.1414 && & 0.0739 & 0.0751 & 0.0756 & 0.0758 & 0.0760 \\  \\

Nat. Cell Biol. & 0.0801 & 0.0030 & 0.0001 & 3.12E-6 & 9.2E-8 && Jour. of Econ. Persp. &0.2337 & 0.0664 & 0.0195 & 0.0058 & 0.0017  \\
                    & 0.1443 & 0.1546 & 0.1587 & 0.1608 & 0.1621 &&  &  0.0624 & 0.0598 & 0.0588 & 0.0582 & 0.0578  \\  \\

Nat. Meth.  & 0.0724 & 0.0039 & 0.0002 & 1.1E-5 & 5.43E-7 && Jour. of Finan. Econ.  &0.2267 & 0.0322 & 0.0040 & 0.0004 & 5.7E-5     \\  
                & 0.1328 & 0.1363 & 0.1376 & 0.1383 & 0.1388  &&  & 0.0885 & 0.0963 & 0.0995 & 0.1012 & 0.1024 \\  \\

Plant C. & 0.1330 & 0.0081 & 0.0004 & 1.9E-5 & 8.70E-7 && Quart. Jour. of Econ. & 0.2237 & 0.0453 & 0.0090 & 0.0017 & 0.0003 \\
           & 0.1234 & 0.1356 & 0.1405 & 0.1431 & 0.1447  &&  & 0.0758 & 0.0773 & 0.0779 & 0.0782 & 0.0784 \\  \\

Plant Jour. & 0.1401 & 0.0098 & 0.0005 & 3.2E-5 & 1.73E-6 && Small Bus. Econ. & 0.2108 & 0.0242 & 0.0024 & 0.0002 & 1.9E-5   \\
              & 0.1182 & 0.1291 & 0.1334 & 0.1357 & 0.1372  &&  & 0.0966 & 0.1068 & 0.1110 & 0.1133 & 0.1148   \\  \\

IEEE Comm. Mag. & 0.0140 & 0.0001 & 1.19E-6 & 1.04E-8 & 8.8E-11 && Acc. of Chem. Res. & 0.1065 & 0.0103 & 0.0009 & 9.2E-5 & 8.6E-6     \\
         & 0.2052 & 0.2093 & 0.2108 & 0.2116 & 0.2121 && & 0.1095 & 0.1107 & 0.1112 & 0.1115 & 0.1116  \\  \\

IEEE Sens. Jour. & 0.1001 & 0.0071 & 0.0004 & 3.0E-5 & 1.9E-6 && ACS Catalysis  & 0.1454 & 0.0184 & 0.0022 & 0.0002 & 3.3E-5    \\  
         & 0.1212 & 0.1258 & 0.1277 & 0.1286 & 0.1293 && & 0.0969 & 0.0987 & 0.0995 & 0.0999 & 0.1001   \\   \\

IEEE Trans. Cyb. & 0.0963 & 0.0045 & 0.0001 & 7.18E-6 & 2.6E-6  && Adv. Funct. Mat.     &0.0640 & 0.0033 & 0.0001 & 8.18E-6 & 3.9E-7        \\
         & 0.1353 & 0.1455 & 0.1495 & 0.1516 & 0.1529 &&  & 0.1359 & 0.1385 & 0.1395 & 0.1400 & 0.1404   \\  \\

IEEE Trans. VT  & 0.0593 & 0.0025 & 0.0001 & 4.08E-6 & 1.6E-7  && Chem   & 0.1472 & 0.0209 & 0.0029 & 0.0004 & 5.8E-5         \\
 &  0.1432 & 0.1472 & 0.1488 & 0.149625 & 0.1501 && & 0.0926 & 0.0930 & 0.0932 & 0.0933 & 0.0934     \\  \\

IEEE Wir. Comm. & 0.0046 & 9.2E-6 & 1.6E-8 & 2.6E-11 & 4.03E-14 &&  Chem. Record & 0.1239 & 0.0088 & 0.0005 & 3.3E-5 & 1.9E-6       \\
 &  0.2625 & 0.2708 & 0.2739 & 0.2755 & 0.2764 && & 0.1190 & 0.12719 & 0.1304 & 0.1321 & 0.1332        \\  \\

Nano En.  & 0.0435 & 0.0016 & 5.9E-5 & 2.16E-6 & 7.7E-8   && Chem. Science                 & 0.1498 & 0.0197 & 0.0025 & 0.0003  & 4.1E-5            \\
 & 0.1503 & 0.1521 & 0.1527 & 0.1531 & 0.1533 && & 0.0952 & 0.0970 & 0.0977 & 0.0980 & 0.0983 \\  \\

Nat. Electr.  &  0.0791 & 0.0049 & 0.0002 & 1.7E-5 & 1.00E-6 && Chem. of Mat.  & 0.1287  & 0.0159 & 0.0019 & 0.0002 & 2.9E-5     \\
                    &  0.1277 & 0.1307 & 0.1319 & 0.1325 & 0.1329 &&  & 0.0987 & 0.0992 & 0.0994 & 0.0995 & 0.0995        \\  \\

Nat. Nanot.  & 0.0967 & 0.0072 & 0.0005 & 3.5E-5 & 2.4E-6 && Jour. of Mat. Chem. B & 0.0619 & 0.0025 & 9.4E-5 & 3.4E-6 & 1.2E-7                \\
 &  0.1197 & 0.1231 & 0.1245 & 0.1252 & 0.1257 && & 0.1449 & 0.1502 & 0.1523 & 0.1534 & 0.1541       \\  \\

Procc. IEEE  & 0.1600 & 0.0282 & 0.0050 & 0.0009 & 0.0001 && Nano Letters  & 0.1077 & 0.0097 & 0.0008 & 7.3E-5 & 6.2E-6                   \\
  & 0.0837 & 0.0824 & 0.0819 & 0.0816 & 0.0814 && & 0.1119 & 0.1142& 0.1151 & 0.1156 & 0.115986   \\  \\

Sig. Process.      & 0.1394 & 0.0132 & 0.0011 & 9.7E-5 & 8.01E-6 && Nat. Mater.  &0.1207 & 0.0112 & 0.0009 & 8.6E-5 & 7.3E-6                  \\
 &  0.1078 & 0.1134 & 0.1156 & 0.1168 & 0.1176 && & 0.1098 & 0.1133 & 0.1147 & 0.1155 & 0.1159    \\
      \hline

\end{longtable}
\end{center}
\end{landscape}}

{\tiny
\begin{landscape}
\begin{center}
\begin{longtable}{lrrrrrrlrrrrr}

 \caption{{From top to down survival rate and mortality rate for journals in GM, GP\&A, H, and L\&IS}\label{tabA4}}

\\
\hline
& \multicolumn{5}{c}{$Age$}  &&& \multicolumn{5}{c}{$Age$} \\ \cline{2-6}\cline{9-13}
& 20 & 40 & 60 & 80 & 100  &&& 20 & 40 & 60 & 80 & 100\\ \cline{2-6}\cline{9-13}
\hline
\endfirsthead
\multicolumn{13}{c}%
{\tablename\ \thetable\ -- \textit{Continued from previous page}} \\
\hline
& \multicolumn{5}{c}{$Age$}  &&& \multicolumn{5}{c}{$Age$} \\ \cline{2-6}\cline{9-13}
& 20 & 40 & 60 & 80 & 100  &&& 20 & 40 & 60 & 80 & 100\\ \cline{2-6}\cline{9-13}
\hline
\endhead
\hline \multicolumn{13}{r}{\textit{Continued on next page}} \\
\endfoot
\hline
\endlastfoot

BMC Med. & 0.0811 & 0.0040 & 0.0001 & 8.1E-6 & 3.5E-7&& Adv. in Theor. \& MP  &   0.3421 & 0.0927 & 0.0237 & 0.0059 & 0.0014 \\
 & 0.1351  & 0.1416 & 0.1441 & 0.1454 & 0.1462 && & 0.0608 & 0.0649 & 0.0666 & 0.0676 & 0.0682 \\ \\

Clin. Sc. & 0.0803 & 0.0033 & 0.0001 & 4.2E-6 & 1.42E-7 && Chaos & 0.2241 & 0.0468 & 0.0096 & 0.0019 & 0.0004 \\
 & 0.1414 & 0.1504 & 0.1539 & 0.1557 & 0.1569  && & 0.0747 & 0.0757 & 0.0761 & 0.0763 & 0.0765 \\ \\

Front. Med. & 0.1215 & 0.0110 & 0.0009 & 7.9E-5 & 6.5E-6 && Comb. and Flame & 0.2450 & 0.0537 & 0.0114 & 0.0024 & 0.0005 \\
 & 0.1106 & 0.1146 & 0.1162 & 0.11715 & 0.117694  &&  & 0.0721 & 0.0738 & 0.0745 & 0.0749 & 0.0751 \\ \\

JAMA-Jour. AMA &  0.0904 & 0.0059 & 0.0003 & 2.2E-5 & 1.3E-6 && Comm. in Number T\&P & 0.3249 & 0.0985 & 0.0294 & 0.0087 & 0.0025 \\
 &  0.1248 & 0.1290 & 0.1307 & 0.1316 & 0.1321    &&   &   0.0572 & 0.0583 & 0.0588 & 0.0590 & 0.0592 \\ \\

Jour. Clin. Inv. &  0.1003 & 0.0047 & 0.0001 & 7.1E-6 & 2.5E-7  && Comp. Physics Commun. &      0.2745 & 0.0604 & 0.0126 & 0.0025 & 0.0005 \\
 &  0.1351 & 0.1462 & 0.1506 & 0.1529 & 0.1543 && &  0.0708 & 0.0744 & 0.0759 & 0.0767 & 0.0773 \\ \\

Jour.  Exp. Med. &  0.1013 & 0.0050 & 0.0002 & 8.7E-6 & 3.3E-7 &&  Jour. of Biophot. &   0.1260 & 0.0095 & 0.0006 & 4.1E-5 & 2.5E-6 \\
 &  0.1331 & 0.1435 & 0.1476 & 0.1498 & 0.1511 && & 0.1169 & 0.1245 & 0.1275 & 0.1291 & 0.1301 \\ \\

New Eng. Jour. Med. &  0.0956 & 0.0076 & 0.0005 & 4.4E-5 & 3.3E-6   &&  Jour. of Comp. Physics & 0.2565 & 0.0534 & 0.0106 & 0.0020 & 0.0003 \\
 &  0.1174 & 0.1198 & 0.1207 & 0.1212 & 0.1215 && & 0.0735 & 0.0769 & 0.0782 & 0.0790 & 0.0795 \\ \\

PLoS Medicine &  0.0639 & 0.0025 & 9.0E-5 & 3.1E-6 & 1.03E-7&&   Jour. of Physical \& CRD &  0.5090 & 0.2391 & 0.1100 & 0.0501 & 0.0226 \\
 &  0.1458 & 0.1521 & 0.1545 & 0.1558 & 0.1567 && &  0.0362 & 0.0377 & 0.0383 & 0.0387 & 0.0389 \\ \\

The BMJ &  0.0842 & 0.0059 & 0.0004 & 2.7E-5 & 1.87E-6 && Jour. of Phys. A: M\&T    &0.3199 & 0.0973 & 0.0293 & 0.0087 & 0.0026 \\
 &  0.1226 & 0.1248 & 0.1257 & 0.1261 & 0.1264  && & 0.0573 & 0.0580 & 0.0584 & 0.0585 & 0.0587 \\ \\

The Lancet &  0.0801 & 0.0050 & 0.0003 & 1.8E-5 & 1.1E-6 && Nat. Phys. & 0.1962 & 0.0347 & 0.0060 & 0.0010 & 0.0001 \\
 &  0.1270 & 0.1300 & 0.1311 & 0.1318 & 0.1321 &&  &  0.0821 & 0.0835 & 0.0841 & 0.0845 & 0.0847 \\ \\

Amer. Ant. &  0.4173 & 0.1588 & 0.0591 & 0.0217 & 0.0079   &&  IEEE Trans. on IT & 0.2678 & 0.0640 & 0.0149 & 0.0034 & 0.0007 \\
 &   0.0462 & 0.0478 & 0.0485 & 0.0489 & 0.0491 && & 0.0680 & 0.0697 & 0.0704 & 0.0708 & 0.0711 \\ \\

Asian Stud. Rev. &   0.2748& 0.0582 & 0.0116 & 0.0022 & 0.0004 &&  Infor. Sys. Man. &  0.1849 & 0.0160 & 0.0011 & 7.5E-5 & 4.7E-6 \\
 &   0.0722 & 0.0764 & 0.0782 & 0.0792 & 0.0798 &&   & 0.1077 & 0.1206  & 0.1259 & 0.1287 & 0.1304 \\ \\

Hist. and Anthrop. &   0.4393 & 0.1921 & 0.0839 & 0.0366 & 0.0160 &&  Infor. Syst. Res.  &     0.2415 & 0.0303 & 0.0032 & 0.0003 & 2.9E-5 \\
 &   0.0404 & 0.0405 & 0.0405 & 0.0405 & 0.0405  && & 0.0918 & 0.1036 & 0.1084 & 0.1110 & 0.1127 \\ \\

Hist. of the Human Sc. &   0.5370 & 0.2840 & 0.1496 & 0.0786 & 0.0413 &&  Int. Jour. of IM  &    0.1284 & 0.0111 & 0.0008 & 6.8E-5 & 5.2E-6 \\
 &   0.0311 & 0.0314 & 0.0315 & 0.0316 & 0.0317 && & 0.1118 & 0.1174 & 0.1196 & 0.1208 & 0.1216 \\ \\

Jour. of East. Afr. Stud. &   0.2155 & 0.0299 & 0.0037 & 0.0004 & 5.1E-5   &&  Jour. of Infor. Sc.  &   0.0897 & 0.0035 & 0.0001 & 3.6E-6 & 1.0E-7 \\
 &   0.0899 & 0.0971 & 0.1000 & 0.1016 & 0.1026 && & 0.1422 & 0.1542 & 0.1588 & 0.1613 & 0.1629 \\ \\

Jour. of Philos. of Educ. &   0.4199 & 0.1642 & 0.0631 & 0.0240 & 0.0091 && Jour. of Inform.  &  0.1738 & 0.0238 & 0.0031 & 0.0003 & 5.0E-5 \\
 &   0.0451 & 0.0463 & 0.0469 & 0.0472 & 0.0473  && & 0.0925 & 0.0960 & 0.0974 & 0.0981 & 0.0986 \\ \\

Latin Amer. Antiq. &   0.4270 & 0.1511 & 0.0509 & 0.0167 & 0.0054 && Jour. of the Ass. for ICT  &   0.1858 & 0.0186 & 0.0016 & 0.0001 & 1.0E-5 \\
 &   0.0486 & 0.0521 & 0.0536 & 0.0544 & 0.0550 && &  0.1027 & 0.1128 & 0.1169 & 0.1192 & 0.1206 \\ \\

Past and Present &   0.4686 & 0.1947 & 0.0784 & 0.0311 & 0.0122  &&  Res. Eval. & 0.1458 & 0.0127 & 0.0009 & 7.4E-5 & 5.3E-6 \\
 &   0.0417 & 0.0439 & 0.0448 & 0.0454 & 0.0458  && &  0.1102 & 0.1179 & 0.1210 & 0.1227 & 0.1237 \\ \\

Philos. of Sc. &   0.3684 & 0.1245 & 0.0412 & 0.0135 & 0.0044 && Scientom. & 0.1526 & 0.0188 & 0.0022 & 0.0002 & 2.9E-5 \\
 &   0.0519 & 0.0534 & 0.0540 & 0.0543 & 0.0545 &&  &  0.0976 & 0.1006 & 0.1018 & 0.1024 & 0.1029 \\ \\

Stud. in H\&O of Science Part A  &   0.4491 & 0.1861 & 0.0756 & 0.0304 & 0.0121 &&  Social Sc. Comp. Rev. & 0.1451 & 0.0156 & 0.0015 & 0.0001 & 1.5E-5 \\
 &   0.0422 & 0.0437 & 0.0443 & 0.0446 & 0.0449 && & 0.1030 & 0.1073 & 0.1090 & 0.1099 & 0.1105 \\

\end{longtable}
\end{center}
\end{landscape}}

{\fontsize{4}{5}\selectfont
\begin{landscape}
\begin{center}
\begin{longtable}{lrrrrrrrrrrrlrrrrrrrrr}

 \caption{Percentile VaR (above) and TVaR (below) for all the journals considered\label{tabA5}}

\\
\hline
Journal & \multicolumn{9}{c}{$p$} && Journal & \multicolumn{9}{c}{$p$}\\ \cline{2-11} \cline{13-21}
& 0.01 & 0.02 & 0.03 & 0.04 & 0.05 & 0.06 & 0.07 & 0.08 & 0.09 &&

& 0.01 & 0.02 & 0.03 & 0.04 & 0.05 & 0.06 & 0.07 & 0.08 & 0.09\\ \cline{2-11} \cline{13-21}
\hline
\endfirsthead
\multicolumn{13}{c}%
{\tablename\ \thetable\ -- \textit{Continued from previous page}} \\
\hline
Journal & \multicolumn{9}{c}{$p$} && Journal & \multicolumn{9}{c}{$p$}\\ \cline{2-11} \cline{13-21}
& 0.01 & 0.02 & 0.03 & 0.04 & 0.05 & 0.06 & 0.07 & 0.08 & 0.09 &&

& 0.01 & 0.02 & 0.03 & 0.04 & 0.05 & 0.06 & 0.07 & 0.08 & 0.09\\ \cline{2-11} \cline{13-21}
\hline
\endhead
\hline \multicolumn{13}{r}{\textit{Continued on next page}} \\
\endfoot
\hline
\endlastfoot

Ag.Cell     &33	&29	&26	&25	&23	&22	&21	&20	&19  &&Amer. Econ. Rev.    &    64	&54	&49	&45	&42	&39	&37	&35	&33  \\
            & 34.22 & 30.22 & 27.22 & 26.22 & 24.22 & 23.22 & 22.22 & 21.22 & 20.22  &&    & 65.07 & 55.07 & 50.07 & 46.07 & 43.07 & 40.07 & 38.07 & 36.07 & 34.07\\
Autoph.               &29	&26	&24	&22	&21	&20	&19	&19	&18  &&Appl. Econ.                      &    49	&43	&39	&37	&34	&33	&31	&30	&29  \\
& 30.30 & 27.30 & 25.30 & 23.30 & 22.30 & 21.30 & 20.31 & 20.31 & 19.31  &&  & 50.13 & 44.13 & 40.13 & 38.13 & 35.13 & 34.13 & 32.13 & 31.13 & 30.13\\
Blood                   &37	&32	&29	&27	&25	&24	&23	&22	&21  &&Appl. Econ. Letters              &    51	&44	&40	&37	&35	&33	&31	&30	&29  \\
& 38.16 & 33.16 & 30.16 & 28.16 & 26.16 & 25.16 & 24.16 & 23.16 & 22.16  &&  & 52.11 & 45.11 & 41.11 & 38.11 & 36.11 & 34.11 & 32.11 & 31.11 & 30.11\\
Canc. C.             &27	&24	&22	&20	&19	&18	&18	&17	&16  &&Ecolog. Econ.                    &    43	&37	&34	&31	&29	&28	&26	&25	&24  \\
& 28.26 & 25.27 & 23.27 & 21.27 & 20.27 & 19.27 & 19.27 & 18.27 & 17.27  && & 44.14 & 38.14 & 35.14 & 32.14 & 30.14 & 29.14 & 27.14 & 26.14 & 25.14\\
C. Metab.         &36	&31	&28	&26	&25	&24	&22	&21	&21  &&Econom. Jour.                    &    55	&48	&44	&41	&39	&37	&35	&34	&32  \\
& 37.17 & 32.17 & 29.17 & 27.17 & 26.17 & 25.17 & 23.17 & 22.17 & 22.17  &&  & 56.11 & 49.11 & 45.11 & 42.11 & 40.11 & 38.11 & 36.11 & 35.11 & 33.11\\
Jour. C. Biol.     &39	&34	&31	&29	&27	&26	&25	&24	&23  &&Econ. Letters                    &    60	&51	&46	&42	&39	&37	&35	&33	&32  \\
& 40.17 & 35.17 & 32.17 & 30.18 & 28.18 & 27.18 & 26.18 & 25.18 & 24.18  && & 61.08 & 52.08 & 47.08 & 43.08 & 40.08 & 38.08 & 36.08 & 34.08 & 33.08\\
Nat. Cell Biol.         &32	&28	&26	&24	&23	&21	&20	&20	&19  &&Jour. of Econ. Persp.            &    71	&59	&52	&48	&44	&41	&39	&36	&35  \\
& 33.20 & 29.20 & 27.20 & 25.20 & 24.20 & 22.20 & 21.20 & 21.20 & 20.21  && & 72.05 & 60.05 & 53.05 & 49.05 & 45.05 & 42.05 & 40.05 & 37.05 & 36.05\\
Nat. Meth.            &33	&28	&26	&24	&22	&21	&20	&19	&18  &&Jour. of Finan. Econ.            &    51	&44	&40	&37	&35	&33	&32	&30	&29  \\
& 34.16 & 29.16 & 27.16 & 25.16 & 23.16 & 22.16 & 21.16 & 20.16 & 19.16  && & 52.12 & 45.12 & 41.12 & 38.12 & 36.12 & 34.12 & 33.12 & 31.12 & 30.12\\
Plant C.              &38	&33	&30	&28	&27	&25	&24	&23	&22  &&Quart. Jour. of Econ.            &    58	&50	&45	&41	&38	&36	&34	&32	&31  \\
& 39.18 & 34.18 & 31.18 & 29.18 & 28.18 & 26.18 & 25.18 & 24.18 & 23.18  && & 59.08 & 51.08 & 46.08 & 42.08 & 39.08 & 37.08 & 35.08 & 33.08 & 32.08\\
Plant Jour.           &39	&34	&31	&29	&27	&26	&25	&24	&23  &&Small Bus. Econ.                 &    47	&41	&38	&35	&33	&31	&30	&29	&28  \\
& 40.17 & 35.17 & 32.17 & 30.17 & 28.17 & 27.17 & 26.17 & 25.17 & 24.17  && & 48.14 & 42.14 & 39.14 & 36.14 & 34.14 & 32.14 & 31.14 & 30.14 & 29.14\\ \\

IEEE Comm. Mag.              &21	&18	&16	&15	&14	&13	&12	&12	&11  && Acc. of Chem. Res.    &  40	&34	&30	&28	&26	&24	&23	&22	&21   \\
& 22.27 & 19.27 & 17.27 & 16.27 & 15.27 & 14.27 & 13.27 & 13.27 & 12.27&& & 41.12 & 35.12 & 31.12 & 29.12 & 27.12 & 25.12 & 24.12 & 23.12 & 22.12 \\
IEEE Sens. Jour.                      &37	&32	&29	&27	&25	&23	&22	&21	&20  && ACS Catalysis                    &  45	&39	&35	&32	&30	&28	&27	&25	&24   \\
& 38.15 & 33.15 & 30.15 & 28.15 & 26.15 & 24.15 & 23.15 & 22.15 & 21.15&& & 46.11 & 40.11 & 36.11 & 33.11 & 31.11 & 29.11 & 28.11 & 26.11 & 25.11 \\
IEEE Trans. Cyb.                    &35	&30	&27	&25	&24	&23	&22	&21	&20  && Adv. Funct. Mat.     &  32	&27	&25	&23	&21	&20	&19	&18	&17   \\
& 36.19 & 31.19 & 28.19 & 26.19 & 25.19 & 24.19 & 23.19 & 22.19 & 21.19&& & 33.16 & 28.16 & 26.16 & 24.16 & 22.16 & 21.16 & 20.16 & 19.16 & 18.16 \\
IEEE Trans. VT                   &31	&26	&24	&22	&21	&19	&18	&18	&17  && Chem                             &  47	&40	&36	&33	&31	&29	&27	&26	&25   \\
& 32.18 & 27.18 & 25.18 & 23.18 & 22.18 & 20.18 & 19.18 & 19.18 & 18.18&& & 48.10 & 41.10 & 37.10 & 34.10 & 32.10 & 30.10 & 28.10 & 27.10 & 26.10 \\
IEEE Wir. Comm.                       &17	&15	&13	&12	&12	&11	&10	&10	&9   && Chem. Record                   &  39	&33	&30	&28	&27	&25	&24	&23	&22   \\
& 18.40 & 16.40 & 14.41 & 13.41 & 13.41 & 12.41 & 11.41 & 11.41 & 10.41&& & 40.16 & 34.16 & 31.16 & 29.16 & 28.16 & 26.16 & 25.16 & 24.16 & 23.16 \\
Nano En.                               &28	&24	&22	&20	&19	&18	&17	&16	&15  && Chem. Science                 &  46	&39	&35	&33	&30	&29	&27	&26	&25   \\
& 29.18 & 25.18 & 23.18 & 21.18 & 20.18 & 19.18 & 18.18 & 17.18 & 16.18&&  & 47.11 & 40.11 & 36.11 & 34.11 & 31.11 & 30.11 & 28.11 & 27.11 & 26.11 \\
Nat. Electr.                              &34	&29	&27	&24	&23	&22	&20	&19	&19  && Chem. of Mat.            &  44	&37	&33	&31	&29	&27	&25	&24	&23   \\
& 35.15 & 30.15 & 28.15 & 25.15 & 24.15 & 23.15 & 21.15 & 20.15 & 20.15&& & 45.11 & 38.11 & 34.11 & 32.11 & 30.11 & 28.11 & 26.11 & 25.11 & 24.11 \\
Nat. Nanot.                     &37	&32	&29	&26	&25	&23	&22	&21	&20  && Jour. of Mat. Chem. B &  31	&27	&24	&22	&21	&20	&19	&18	&17   \\
& 38.14 & 33.14 & 30.14 & 27.14 & 26.14 & 24.14 & 23.14 & 22.14 & 21.14&& & 32.18 & 28.18 & 25.18 & 23.18 & 22.18 & 21.19 & 20.19 & 19.19 & 18.19 \\
Procc. IEEE                   &52	&44	&39	&35	&33	&31	&29	&27	&26  && Nano Letters                      &  39	&34	&30	&28	&26	&24	&23	&22	&21   \\
& 53.08 & 45.08 & 40.08 & 36.08 & 34.08 & 32.08 & 30.08 & 28.08 & 27.08&& & 40.13 & 35.13 & 31.13 & 29.13 & 27.13 & 25.13 & 24.13 & 23.13 & 22.13\\
Sig. Process.                             &42	&36	&33	&30	&28	&27	&25	&24	&23  && Nat. Mater.                       &  40	&35	&31	&29	&27	&25	&24	&23	&22   \\
& 43.13 & 37.13 & 34.13 & 31.13 & 29.13 & 28.13 & 26.13 & 25.14 & 24.14&& & 41.13 & 36.13 & 32.13 & 30.13 & 28.13 & 26.13 & 25.13 & 24.13 & 23.13 \\  \\

BMC Med.                                       &34	&29	&26	&24	&23	&22	&21	&20	&19 &&Adv. in Theor. \& MP     &72	&62	&56	&52	&49	&46	&44	&42	&40       \\
& 35.17 & 30.17 & 27.17 & 25.17 & 24.18 & 23.18 & 22.18 & 21.18 & 20.18&& & 73.07 & 63.07 & 57.07 & 53.07 & 50.07 & 47.07 & 45.07 & 43.07 & 41.07 \\
Clin. Sc.                                          &33	&28	&26	&24	&23	&21	&20	&20	&19 &&Chaos                             &59	&50	&45	&42	&39	&36	&34	&33	&31       \\
& 34.19 & 29.19 & 27.19 & 25.19 & 24.19 & 22.19 & 21.19 & 21.19 & 20.20&& & 60.08 & 51.08 & 46.08 & 43.08 & 40.08 & 37.08 & 35.08 & 34.08 & 32.08 \\
Front. Med.                                     &40	&35	&31	&29	&27	&25	&24	&23	&22 &&Comb. and Flame              &61	&52	&47	&43	&40	&38	&36	&34	&33       \\
& 41.13 & 36.13 & 32.13 & 30.13 & 28.13 & 26.13 & 25.13 & 24.13 & 23.13&& & 62.08 & 53.08 & 48.08 & 44.08 & 41.08 & 39.08 & 37.08 & 35.08 & 34.08 \\
JAMA-Jour. AMA                          &36	&31	&28	&26	&24	&23	&21	&20	&20 &&Comm. in Number T\&P     &77	&66	&59	&54	&51	&48	&45	&43	&41       \\
& 37.15 & 32.15 & 29.15 & 27.15 & 25.15 & 24.15 & 22.15 & 21.15 & 21.15&& & 78.06 & 67.06 & 60.06 & 55.06 & 52.06 & 49.06 & 46.06 & 44.06 & 42.06\\
Jour.  Clin. Inv.                                &35	&30	&28	&26	&24	&23	&22	&21	&20 &&Comp. Physics Commun.   &62	&54	&49	&45	&42	&40	&38	&36	&34       \\
& 36.19 & 31.19 & 29.19 & 27.19 & 25.19 & 24.19 & 23.19 & 22.19 & 21.19&& & 63.08 & 55.08 & 50.08 & 46.08 & 43.08 & 41.08 & 39.08 & 37.08 & 35.08 \\
Jour.  Exp. Med.                                 &35	&31	&28	&26	&24	&23	&22	&21	&20 &&Jour. of Biophot.           &39	&34	&31	&29	&27	&25	&24	&23	&22       \\
& 36.19 & 32.19 & 29.19 & 27.19 & 25.19 & 24.19 & 23.19 & 22.19 & 21.19&& & 40.15 & 35.15 & 32.15 & 30.15 & 28.15 & 26.15 & 25.15 & 24.15 & 23.15 \\
New Eng. Jour. Med.                             &37	&32	&29	&26	&25	&23	&22	&21	&20 &&Jour. of Comp. Physics  &60	&52	&47	&43	&40	&38	&36	&34	&33       \\
& 38.14 & 33.14 & 30.14 & 27.14 & 26.14 & 24.14 & 23.14 & 22.14 & 21.14&& & 61.08 & 53.08 & 48.08 & 44.08 & 41.08 & 39.08 & 37.08 & 35.08 & 34.08 \\
PLoS Medicine                                      &31	&27	&24	&22	&21	&20	&19	&18	&17  &&Jour. of Physical \& CRD       &120& 103	& 92 & 85	&8 0	& 75	& 71	& 68 & 65     \\
& 32.19 & 28.19 & 25.19 & 23.19 & 22.19 & 21.19 & 20.19 & 19.19 & 18.19&& & 121.0 & 104.0 & 93.04 & 86.04 & 81.04 & 76.04 & 72.04 & 69.04 & 66.04\\
The BMJ                                            &36	&30	&27	&25	&23	&22	&21	&20	&19 &&Jour. of Phys. A: M\&T    &   77	&66	&59	&54	&51	&48	&45	&43	&41       \\
& 37.14 & 31.14 & 28.14 & 26.14 & 24.14 & 23.14 & 22.14 & 21.14 & 20.14&& & 78.06 & 67.06 & 60.06 & 55.06 & 52.06 & 49.06 & 46.06 & 44.06 & 42.06 \\
The Lancet                                         &35	&30	&27	&25	&23	&22	&20	&20	&19  &&Nat. Phys.                   &54	&46	&41	&38	&35	&33	&31	&30	&29       \\ \\
& 36.15 & 31.15 & 28.15 & 26.15 & 24.15 & 23.15 & 21.15 & 21.15 & 20.15&& & 55.09 & 47.09 & 42.09 & 39.09 & 36.09 & 34.09 & 32.09 & 31.09 & 30.09 \\

Am. Ant.   & 95	&81	&73	&67	&63	&59	&56	&53	&51 && IEEE Trans. Inf. Th. & 65	&56	&50	&46	&43	&40	&38	&36	&35     \\
& 96.05   & 82.05  & 74.05 & 68.05 & 64.05 & 60.05 & 57.05 & 54.05 & 52.05&& & 66.07 & 57.07 & 51.07 & 47.07 & 44.07 & 41.07 & 39.07 & 37.07 & 36.07 \\
As. Stud. Rev.   & 61	&53	&48	&44	&41	&39	&37	&36	&34 &&Infor. Sys. Man.      &43	&38	&35	&32	&30	&29	&28	&27	&26     \\
& 62.09   & 54.09  & 49.09 & 45.09 & 42.09  & 40.09  & 38.09 & 37.09 & 35.09&& & 44.16 & 39.16 & 36.16 & 33.16 & 31.16 & 30.16 & 29.16 & 28.16 & 27.16 \\
Hist. Antrop.                                  & 111   &	94&	84&	77&	72&	68&	64&	61&	58&& Inf. Syst. Res.  & 50&	43&	40&	37&	35&	33&	32&	30&	29    \\
& 112.04  & 95.04  & 85.04 & 78.04 & 73.04 & 69.04 & 65.04 & 62.04 & 59.04&&  & 51.13 & 44.13 & 41.13 & 38.14 & 36.14 & 34.14 & 33.14 & 31.14 & 30.14 \\
Hist. the Human Sc.  & 143   &	12&2	1&09	&101&	94&	88&	83&	79&&Int. Jour. Inf. Manag.    &	 40&	35&	32&	29&	27&	26&	25&	23 & 22\\
& 144.03  & 123.03 & 110.03 & 102.03 & 95.03  & 89.03  & 84.03  & 80.03  & 76.03 &&  & 41.14 & 36.14 & 33.14 & 30.14 & 28.14 & 27.14 & 26.14 & 24.14 & 23.14 \\
Jour. East. Afr. Stud. & 50	&43	&39	&37	&34	&33	&31	&30	&29 && Jour. Inf. Sc.   &33	&29	&26	&25	&23	&22	&21	&20	&19     \\
& 51.12   & 44.12  & 40.12 & 38.12 & 35.12 & 34.12 & 32.12 & 31.12 & 30.12&& & 34.21 & 30.21 & 27.21 & 26.21 & 24.21 & 23.21 & 22.21 & 21.21 & 20.21\\
Jour. of Philos. of Educ.                           & 98	&83	&75	&69	&64	&61	&57	&55	&52 && Jour. Inform.   &48	&41	&37	&34	&32	&30	&29	&27	&26     \\
& 99.050  & 84.05  & 76.05 & 70.05 & 65.05 & 62.05 & 58.050 & 56.05 & 53.05&& & 49.11 & 42.11 & 38.11 & 35.11 & 33.11 & 31.11 & 30.11 & 28.11 & 27.11 \\
Latin Amer. Antiq.   & 89	&76	&69	&64	&60	&57	&54	&51	&49 &&Jour. of the Ass. for ICT  &45	&39	&36	&33	&31	&30	&28	&27	&26     \\
& 90.06   & 77.06  & 70.06 & 65.06 & 61.06 & 58.06 & 55.06 & 52.06 & 50.06&& & 46.14 & 40.14 & 37.14 & 34.14 & 32.14 & 31.14 & 29.14 & 28.14 & 27.14 \\
Past. Pres.  & 104&	89&	80&	74&	69&	65&	62&	59&	57&&Res. Eval.                 &41	&36	&33	&30	&28	&27	&26	&25	&24     \\
& 105.05  & 90.05  & 81.05 & 75.05 & 70.05 & 66.05 & 63.05 & 60.05 & 58.05&& & 42.14 & 37.14 & 34.15 & 31.15 & 29.15 & 28.15 & 27.15 & 26.15 & 25.15 \\
Philos. Sc.   & 85	&73	&65	&60	&56	&53	&50	&48	&45 && Scient. &45	&39	&35	&32	&30	&29	&27	&26	&25     \\
& 86.05   & 74.05  & 66.05  & 61.05  & 57.05  & 54.05  & 51.05 & 49.05 & 46.05&& & 46.11 & 40.11 & 36.11 & 33.11 & 31.11 & 30.11 & 28.11 & 27.11 & 26.11 \\
Stud. H\&O Sc. A   & 104  &89&	80&	74&	69&	65&	61&	58 & 56 && Soc. Sc. Comp. Rev.  & 43	& 37	& 34	& 31	& 29	& 28	& 26	& 25	& 24     \\
& 105.04  & 90.04  & 81.04 & 75.048 & 70.04 & 66.04 & 62.04 & 59.04 & 57.04&& & 44.12 & 38.12 & 35.12 & 32.12 & 30.12 & 29.12 & 27.12 & 26.12 & 25.12 \\ \hline

\end{longtable}
\end{center}
\end{landscape}}

\end{document}